\documentclass[usenatbib]{mnras}
\usepackage{graphicx}
\usepackage{hyperref}
\def\apjs{ApJS}
\def\apjl{ApJL}
\def\mnras{MNRAS}
\def\aap{A\&A}

\usepackage{amsmath,color}
\usepackage{deluxetable}
\usepackage{color}
\usepackage{soul}
\usepackage{xcolor}
\usepackage{dblfloatfix}

\title[Functions of Red W1-W2 Galaxies]{The Luminosity and Stellar Mass Functions of Red W1-W2 Galaxies}
\author[O'Connor et al.]{J. A. O'Connor,$^{1}$\thanks{E-mail: joconno5@gmu.edu}
 J. L. Rosenberg$^{1}$
 S. Satyapal$^{1}$
 N.J. Secrest$^{1,2}$\\
$^{1}$Department of Physics and Astronomy, George Mason University, Fairfax, VA 22030\\
$^{2}$United States Naval Observatory\\
}

\date{Accepted 2016 August 05. Recieved 2016 July 24; in original form 2015 December 16}

\pubyear{2016} 

\begin{document}
\label{fistpage}
\pagerange{\pageref{firstpage}--\pageref{lastpage}}
\maketitle

\begin{abstract}

We present a study of nearby galaxies as a function of their [3.4]-[4.6] colour.
Galaxies that are red in their [3.4]-[4.6] colour contain heated dust and the reddest
systems ([3.4]-[4.6] $>$ 0.5) are classified as AGN by some selection criteria. The sample discussed here includes nearby galaxies selected from the Sloan Digital Sky Survey (SDSS) that are also in the Wide-field Infrared Survey Explorer (WISE) catalogue. We
calculate the number density of galaxies, in the form of the luminosity and mass functions, using the V/Vmax method and a Stepwise Maximum Likelihood method that has been modified to account for the additional colour selection. The reddest galaxies which have [3.4]-[4.6] $>$ 0.8 and are sometimes classified as AGN by their colour, make up 0.2\% of nearby galaxies. However, the reddest galaxies are a rising fraction of the low mass galaxy population. Identifying the lowest mass (M $<$ 10$^{8}$M$_{\odot}$) red ([3.4]-[4.6] $>$ 0.8) galaxies as AGN is surprising given that none are optical AGN or composites, in contrast with their more massive (M $>$ 10$^{10}$M$_{\odot}$) red galaxy counterparts that are dominated by optical AGN and composites (86.4\%). We also show that these low mass red galaxies are associated with higher specific star formation rates than their bluer counterparts. While the properties of this relatively rare segment of nearby low-mass galaxies are intriguing, particularly if they are associated with AGN activity, there is not yet enough evidence to determine whether it is AGN or unusual star formation that is driving red colours in these systems.

\end{abstract}


\section{Introduction}

Infrared (IR) colours of galaxies observed by the \textit{Wide-Field Infrared
Survey Explorer} (WISE) provide a probe of hot dust emission
\citep{Jarrett2011,Donley2012,Mateos2012,Stern2012}. Active galactic nuclei (AGN) activity and star formation are
the dominant processes that heat the dust and thus galaxies with very red 
[3.4]-[4.6] (W1 - W2) colours provide a probe of the most extreme interstellar medium
environments. The luminosity and mass functions of galaxies with red W1 - W2
colours measure the prevalence of these extreme environments in the local universe.

AGNs have a hard radiation field that, in dusty systems, is absorbed and
re-radiated in the infrared. The spectral energy distributions (SEDs) of AGNs are
predicted to display power law slopes ($f_{\nu}\ \propto\ \nu^{\alpha}$ with $\alpha\ <\ -0.5$)
 \citep{AlonsoHerrero2006} in the infrared producing red colours in W1 - W2 (or
equivalently the [3.6] - [4.5] Spitzer bands;
\citealp{Richards2006,Polletta2007,Assef2010}). Observationally \citet{Yan2013}
have shown that QSOs and some Seyfert galaxies selected from the \textit{Sloan
Digital Sky Survey} (SDSS) have redder W1 - W2 colours than star-forming
galaxies. \citet{Jarrett2011}, \citet{Stern2012}, and \citet{Mateos2012} have
created AGN selection criteria based primarily on a galaxy's W1 - W2 colour,
though \citet{Jarrett2011} and \citet{Mateos2012} also make use of W2 - W3
([4.6] - [12]) colour in their diagnostics. For W1 - W2 $>$ 0.8, 95\% of the
galaxies in the COSMOS field are AGNs \citep{Stern2012}. 
This selection technique is particularly sensitive to heavily
obscured or Compton-Thick AGN, which may not be visible in optical or X-ray
wavelengths \citep{Goulding2011,Assef2013,Mateos2013,Rovilos2014,Stern2014}.

While the aforementioned work shows that AGNs can have extreme MIR colours, 
extreme star formation may also heat the dust in galaxies and lead 
to excess emission in the longer wavlelengths of \textit{WISE}
\citep{Schaerer1997,Ranalli2003,Charmandaris2008}. Longer mid-infrared
wavelengths such as MIPS 24\micron \citep{Kennicutt1998}, 
W3 (12\micron) and W4 (22\micron) \citep{Jarrett2013,Lee2013}, and IRAC 4
(8\micron) \citep{Calzetti2007,Bendo2008} are used for measurments of star
formation rates for this reason. 

\citet{Satyapal2014} found that the fraction of galaxies in the local universe
with extreme mid-infrared colours is higher in low mass galaxies than at high
stellar masses. It is unclear why the number of red low mass galaxies is so
large and whether the dominant cause of their dust heating is extreme star
formation or AGN activity. We examine the number density of these systems and
their contribution to the overall galaxy population and properties from the
optical spectra which may explain the nature of their nuclear activity.
Throughout this paper we will use H$_{0}$=70 km s$^{-1}$ Mpc$^{-1}$, 
$\Omega_{m}$=0.3 and $\Omega_{\Lambda}$=0.7.

\section{Sample Selection}

Our galaxy sample is selected from the Sloan Digital Sky Survey (SDSS)
data release 7 (DR7) catalogue \citep{Abazajian2009} and includes all
spectroscopic objects in the Legacy area with SpecPhoto.specClass=2 (Galaxies)
or SpecPhoto.specClass=3 (QSOs), 0.005 $\leq$ z $\leq$ 0.1 and the Petrosian
$r$-band magnitude, corrected for Galactic 
extinction \citep{Schlegel1998},  
14.0 $\leq$ $r$ $\leq$ 17.77. The SDSS Legacy Survey is a spectroscopic and 
photometric optical survey covering over 8000 deg$^{-2}$ on the sky. 
The faint magnitude limit, $r$=17.77, represents 
the completeness limit of the DR7 spectroscopic sample \citep{Strauss2002}. On 
the other hand, galaxies with $r$ $<$ 14 may have unreliable photometry due 
to shredding \citep{Strauss2002}. The
resulting sample contains $\approx$315,000 galaxies. 

The SDSS sample was matched, within 3$\arcsec$, with galaxies in the AllWISE
Catalogue
\citep{Wright2010,Mainzer2011,Cutri2013}\footnote[1]{http://irsa.ipac.caltech.edu/Missions/wise.html}
that have S/N $>$ 5.0 at both 3.4 $\micron$ (W1) and 4.6 $\micron$ (W2). The
angular resolution of WISE is approximately 6$\arcsec$ in the W1 and W2 bands.
Out of the SDSS galaxies with WISE matches (315251 galaxies or 99.7\% of the
SDSS sample), 6837 (2.2\%) have two WISE matches and 67 (0.002\%) have 3 WISE
matches. In the infrared and optical images, these multiple matches appear to 
be single galaxies that WISE identifies two or three times, generally with the
WISE position straddling the centre of the galaxy, although it is not clear why
in these limited cases WISE separated the galaxies into two or three infrared
sources. For these systems the photometry for the source closest
to the SDSS spectrum was used. Only 0.8\% of galaxies with W1 -W2 $>$ 0.3
have multiple WISE sources where the closest source to the SDSS spectrum is not
also the brightest. For galaxies that are resolved in WISE, the difference in
resolution can cause an underestimation of the W1 band profile flux. Therefore,
in the case of sources with w1rchi2 $>$ 2 we use 11$\arcsec$ aperture photometry
in the W1 and W2 bands instead of the profile magnitudes. 1.4\% of randomly
selected positions within the SDSS footprint (for a selection of 4400 sources)
have a match in the ALLWISE catalogue within 3$\arcsec$ that has W1 and W2 S/N
$\geq$ 5.0 so almost all of the matches are expected to be legitimate.

Stellar masses for the galaxies were obtained from the NYU-VAGC
\citep{Blanton2005a,BlantonRoweis2007}\footnote[2]{http://sdss.physics.nyu.edu/vagc/}
which were derived from principal component analysis fits to stellar templates. 
Emission line fluxes used to calculate star formation rates
\citep{Brinchmann2004} and metallicities \citep{Tremonti2004} were obtained from
the MPA-JHU catalogue\footnote[3]{http://home.strw.leidenuniv.nl/~jarle/SDSS/}. 
This catalogue contains line fluxes corrected for stellar absorption using the
\citet{Bruzual2003} stellar population templates.

\section{Methods}

We compute the luminosity and stellar mass functions for the galaxies in our sample
using the 1/$V_{Max}$ \citep{Schmidt1968} method 
and a modified form of the stepwise maximum likelihood method (SWML;
\citealp{Efstathiou1988}) which are not susceptible to the same biases
and therefore provide a check on the results.

The 1/V$_{Max}$ method is used to calculate the number of galaxies per luminosity,
absolute magnitude, or stellar mass bin weighted according to the volume in
which each galaxy could have been detected. For each galaxy in the sample there
are both a minimum and a maximum distance at which it would be included in the V$_{max}$ volume. 
The lower bound to the distance is the larger value of 21 Mpc (defined by the
z=0.005 minimum redshift for the sample) or the minimum distance at which the
galaxy would be included in the sample based on the bright limit $r > 14.0$ for
galaxies in the sample. The maximum comoving distance is 418 Mpc (defined by the redshift
limit of the sample z=0.1) or the maximum distance based on the faint limit,
$r<17.77$ for galaxies in the sample. 

\begin{figure}
\includegraphics[width=\columnwidth]{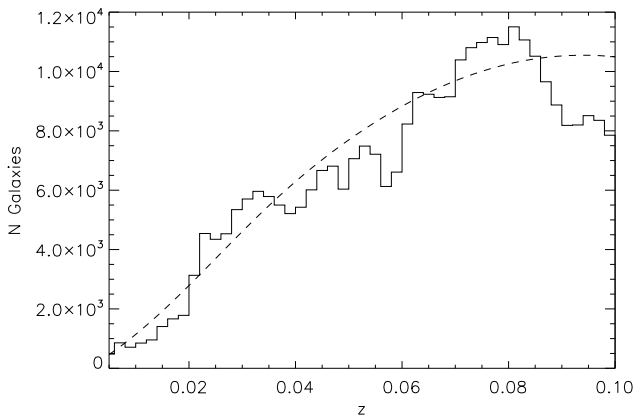}
\caption{The cumulative number of galaxies as a function of redshift in this sample (solid histogram). The dashed line shows the cumulative 
number of galaxies expected as a function of redshift assuming the SDSS DR2 luminosity function \citep{Blanton2005b}.}
\label{Fig:LSS}
\end{figure}

The 1/V$_{Max}$ method has the advantage of being a relatively simple
calculation with built-in normalization, but it relies on 
the assumption that galaxies are distributed homogeneously throughout the survey
volume. The value of $<\frac{V}{V_{Max}}>$ for this sample is 0.4877$\pm$0.0005
where a value of 0.5 indicates an evenly distributed sample.   
The CfA Great Wall \citep{Geller1989} and the Coma Cluster \citep{Harrison2010}
contribute to an excess of galaxies at 0.02 $<$ z $<$ 0.04  while the SDSS Great
Wall \citep{Gott2005} is responsible for an 
excess near z$\approx$0.08.  To mitigate the impact of large-scale seen in Figure \ref{Fig:LSS} structure on
the number density the 1/V$_{Max}$ values have been corrected by a factor of n,
the number of galaxies in the SDSS sample that are within a given redshift range
divided by the number of galaxies expected to fall in that range if they are
distributed according to the \citet{Blanton2005b} $r$-band luminosity function. 

\subsection{Modified Stepwise Maximum Likelihood Method}

The stepwise maximum likelihood method (SWML) \citep{Efstathiou1988} for
measuring the luminosity and mass functions of galaxies has the advantage of
being insensitive to cosmic variance within
the survey volume. For the purposes of this work we modify the standard SWML
technique to account for a galaxy selection (W1-W2 colour) that does
not affect the detectability of the galaxy.

In the standard formulation of the SWML method, the number density of galaxies
in a given bin of luminosity (or absolute magnitude or stellar mass) is given by:

\begin{equation}
\phi_{k}=N_{k}  \Bigg[ \sum_{i=1}^{N_{gal}}\frac{H_{ki}\Delta M_{k}}{\sum_{n=1}^{N_{M}}\phi_{n} \Delta M_{k} H_{ni}} \Bigg]^{-1}
\label{Eq:SWMLmag}
\end{equation}

We derive a normalization for the SWML points by assuming that the total galaxy
density within the luminosity range sampled will match that 
calculated using the 1/V$_{Max}$ method. $N_{k}$ is the number of galaxies in
magnitude bin $k$, $H_{ki}$ is the fraction of bin $k$ 
in which galaxy $i$ is detectable, $N_{gal}$ and $N_{M}$ are the number of
galaxies in the sample and number of absolute magnitude 
bins, respectively. Equation \ref{Eq:SWMLmag} is iterated upon until all
$\phi_{k}$ values change by less than 1\% between successive iterations. 
Equation \ref{Eq:SWMLmass} shows the comparable expression for the mass function
\citep{Loveday2000}. In this case the expression is two-dimensional.
$N_{jk}$ is the number of galaxies in absolute magnitude bin $j$ and stellar
mass bin $k$ and $H_{ji}$ and $H_{ki}$ are the 
fraction of magnitude bin $j$ and stellar mass bin $k$ in which galaxy $i$ could
have been detected.

\begin{equation}
\phi_{jk}=N_{jk} \Bigg[  \sum_{i=1}^{N_{gal}} \frac{H_{ji} H_{ki} \Delta M_{j} \Delta M_{*k}}{\sum_{m=1}^{N_{M}} \sum_{n=1}^{N_{M_{*}}} \phi_{mn} H_{mi} H_{ni} \Delta M_{m} \Delta M_{*n}} \Bigg]^{-1}
\label{Eq:SWMLmass}
\end{equation}

Because the only limiting quantity is the $r$-band apparent magnitude, Equation
\ref{Eq:SWMLmass} can be simplified by assuming that if galaxy i is detectable
in magnitude bin j ($H_{ji} > 0$) then $H_{ki}=1$:

\begin{equation}
\phi_{jk}=N_{jk} \Bigg[ \sum_{i=1}^{N_{gal}} \frac{H_{ji} \Delta M_{j} \Delta M_{*k}}{\sum_{m=1}^{N_{M}} \sum_{n=1}^{N_{M_{*}}} \phi_{mn} H_{mi} \Delta M_{m} \Delta M_{*n}} \Bigg]^{-1}
\label{Eq:SWMLmass2}
\end{equation}

Equation \ref{Eq:SWMLmass2} is then summed over magnitude bins j in order to
calculate values of $\phi_{k}$ \citep{Loveday2000}.

$\phi_{k,x}$ is the number density of galaxies in a given stellar mass
or magnitude bin $k$ and W1-W2 colour bin x. This number density is determined by
multiplying the number density of the full sample by the probability of a galaxy
in bin k falling in colour range x:

\begin{equation}
\phi_{k,x}=\phi_{k} \frac{N_{k,x}}{N_{k}}
\label{Eq:IRphi}
\end{equation}

The 1-sigma uncertainties for the full sample are calculated from the
information matrix \citep{Efstathiou1988}. The inverse of this matrix contains
the variance in $\phi_{k}$. Uncertainties for subsamples are calculated by
adding the uncertainties in $\phi_{k}$ in quaderature with the uncertainties
from Equation \ref{Eq:IRphi}.

\section{Results}\label{results}

\begin{figure*}
\includegraphics[width=\textwidth]{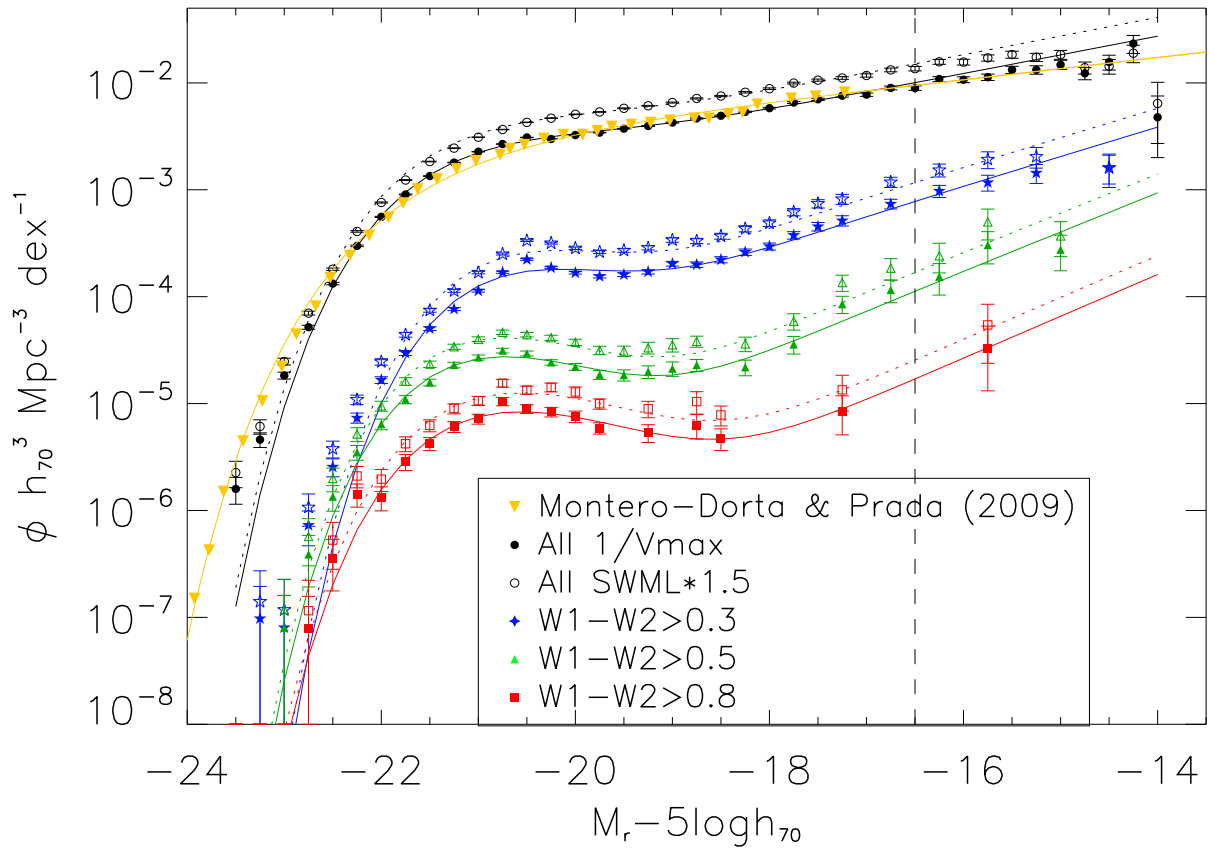}
\caption{The number density of galaxies in our full sample (black circles) and
W1 - W2 colour subsamples (W1 - W2 $\geq$ 0.3, blue stars; W1 - W2$\geq$0.5, green triangles; W1 -
W2$\geq$0.8, red squares) as a function of $r$-band absolute magnitude. 
The figure shows values calculated using the 1/V$_{Max}$ (filled points) and the
SWML (open points). SWML values and errors are multiplied by 1.5 in order to
differentiate them from the 1/V$_{Max}$ points. The plot includes results from
 \citet{Montero2009} (yellow inverted triangles) for comparison.
Lines indicate the best fit Schechter functions. The dashed line at
M$_{r}$=-16.5 is whsere surface brightness effects become important.}
\label{Fig:LSchechter}
\end{figure*}

\begin{figure*}
\includegraphics[width=\textwidth]{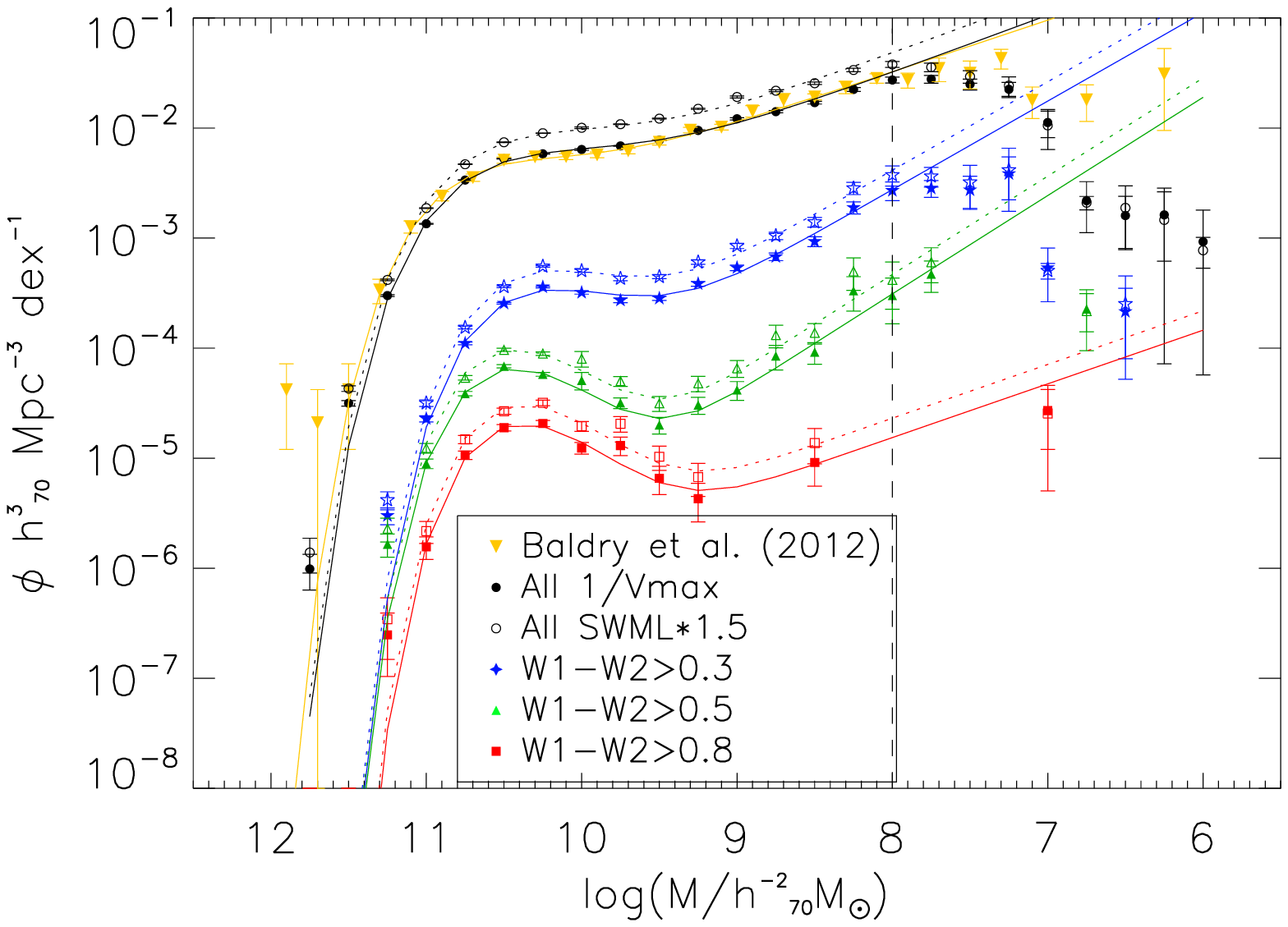}
\caption{The number density of galaxies in our sample and W1 - W2 colour subsamples as a function of stellar mass. Filled points calculated using 
the 1/V$_{Max}$ method and open points are calculated using our modified SWML method. SWML points and errors are shown here multiplied by 1.5 so that they may be more 
easily differentiable from the 1/V$_{Max}$ points. Black circles were calculated using the full galaxy sample while the blue stars only include galaxies with W1 - W2 $\geq $0.3, green triangles only 
include galaxies with W1 - W2 $\geq$ 0.5 and red squares only include galaxies with W1 - W2 $\geq$ 0.8. Solid lines are \citet{Baldry2012}-form Schechter functions fit to the 1/V$_{Max}$ points 
above the surface brightness incompleteness line. The dashed line at 10$^{8}$ M$_{\odot}$ signifies the point below which the mass function is incomplete (See Section \ref{results}). Dotted lines are our best fit Schecter functions multiplied by 1.5 so that their normalization is comparable to 
that of the SWML points. This figure also contains the data points and best fit Schechter function from \citet{Baldry2012} as yellow inverted triangles. }
\label{Fig:MSchechter}
\end{figure*}

Figures \ref{Fig:LSchechter} and \ref{Fig:MSchechter} show the luminosity and
stellar mass functions for galaxies in our sample (black circles) as well as for
subsamples with infrared colours W1 - W2 $\geq$ 0.3 (blue stars), W1 - W2 $\geq$
0.5 (green triangles) and W1 - W2 $\geq$ 0.8 (red squares). In order to distinguish
values determined using the 1/V$_{Max}$ method (filled points, solid lines)
from those determines using the SWML method (open points, dotted lines),
the SWML points have all been multiplied by 1.5. Without this artificial
separation, the results of the two methods lie on top of one another. Solid
lines represent the best fit Schechter functions to the 1/V$_{Max}$ points. The
details of the functions will be discussed later in this section. The
1/V$_{Max}$ and SWML methods yield luminosity and stellar mass functions with
similar shapes indicating that the structure is not due to cosmic variance
within the survey volume.

\par 

The 1/V$_{Max}$ and SWML functions are in good agreement with the yellow points in Figure
\ref{Fig:LSchechter} which show the SDSS DR6 $r$-band luminosity function from
\citet{Montero2009}.
 Figure \ref{Fig:MSchechter} shows the total stellar mass function (black)
plotted alongside the mass function results from \citet{Baldry2012} (yellow).
The downturn at the low mass end of the stellar mass function is due to surface
brightness incompleteness in the SDSS sample and is also seen in
\citet{Baldry2008} and \citet{Baldry2012}. Due to this incompleteness, all
points below log(M/M$_{\odot}$)=8 and M$_{r}$=-16.5 (dashed lines in Figures
\ref{Fig:LSchechter} and \ref{Fig:MSchechter}) are treated as lower limits and
not used to determine the Schechter function fits.

Schecter functions with two characteristic number densities
($\phi_{1}^{*}$, $\phi_{2}^{*}$), one characteristic magnitude or stellar mass
($M^{*}$) and two power law slopes ($\alpha_{1}$, $\alpha_{2}$) were fit to the
points brighter than the surface brightness incompleteness limit. This double
Schecter form is the same one used in \citet{Blanton2005b} and
\citet{Baldry2012}. Parameters of the best fit functions are listed in Tables
\ref{Tab:LFSchechter} and \ref{Tab:MFSchechter}.

The Schecter functions for galaxies with W1 - W2 $\geq$ 0.3, 0.5, and 0.8 have two
components such that the luminosity functions ``dip" at intermediate
luminosities. This dip can be seen in Figure \ref{Fig:LSchechter} and in the
values for $\alpha_{1}$ (the power law slope of the high luminosity population)
$\alpha_{1}$=0.463, $\alpha_{1}$=0.473 and $\alpha_{1}$=0.217 (Table
\ref{Tab:LFSchechter}). This slope is in the opposite direction of the one
measured for the full sample, $\alpha_{1}$=-0.244.
The mass functions in Figure \ref{Fig:MSchechter} and Table
\ref{Tab:MFSchechter} show a similar pattern. For the full stellar mass
function, $\alpha_{1}$=-0.557 indicating a 
smooth transition between the high and low mass galaxy populations. However, the
values of $\alpha_{1}$ for W1 - W2 $\geq$ 0.3, 0.5 and 0.8 are much higher
($\alpha_{1}$=-0.137, 0.423, 0.479). This is evident as a decrease in the number
density from the high to intermediate stellar mass population of red galaxies in
Figure \ref{Fig:MSchechter}.

In addition to this dip, the power law slopes of the red low mass/low luminosity
galaxies are much steeper than those of the total stellar mass and luminosity
functions. For the total stellar mass function, $\alpha_{2}$=-1.524. This
becomes steeper (more negative) for redder galaxy populations. The values of 
$\alpha_{2}$ decrease from -1.804 at W1 - W2 $\geq$ 0.3 to -1.892 at W1 - W2 $\geq$ 0.5
and -1.487 for galaxies with W1 - W2 $\geq$ 0.8. The faint end slopes of the
luminosity functions display a similar trend with values decreasing from
$\alpha_{2}$=-1.438 for the full sample to $\alpha_{2}$=-1.973 for galaxies with
W1-W2$\geq$0.8.
 
The functional forms of the galaxy luminosity and stellar mass functions are
bimodal for W1 - W2 $\geq$ 0.3. The low mass components come close to dominating by number density.
Galaxies with M$_{r}>$ -18 comprise 46.7\% of the total number density (comparing to
galaxies with -24 $<$ M$_{r}$ $<$ -16.5). For galaxies with W1 - W2 $\geq$ 0.5 almost 52\% of
galaxies have M$_{r}$ $>$ -18. However, red galaxies do not make up a large fraction of
the total number density. Galaxies with W1 - W2 $\geq$ 0.3, 0.5 and 0.8 only 
account for 5.6, 0.7 and 0.2\% of the number density of galaxies. 
Low mass galaxies (log(M/M$_{\odot}$) $ < $9) make up 69.4\% of galaxies (8
$<$log(M/M$_{\odot}) <$ 12) with W1 - W2 $\geq$ 0.3, 62.9\% with W1 - W2 $\geq$ 0.5 and
29.9\% with W1 - W2 $\geq$ 0.8. 

Increasing the S/N selection threshold from 5.0 to
10.0 in the W1 and W2 bands for inclusion in our galaxy sample results in a
decrease in the number density of galaxies at low masses and luminosities but
due to the compact nature of the red WISE sources (discussed further in Section
5.1) this mainly affects the samples with W1 - W2 $<$ 0.5. In addition, using the IRAF
task ELLIPSE to calculate the photometry for 60 galaxies spanning a range of
W1-W2 colour, S/N and angular size and comparing them to the values in the
ALLWISE catalogue shows no significant trends in the catalogue colours as a function
of any of these properties. The average offset between the W1-W2 colours from
the AllWISE catalogue and the IRAF photometry is 0.005 mag. In short, there is
no evidence that the red colours of these galaxies are artifacts of poor
photometry.

\subsection{Optical Nuclear Activity Classification}

\begin{figure*}
\includegraphics[width=\textwidth]{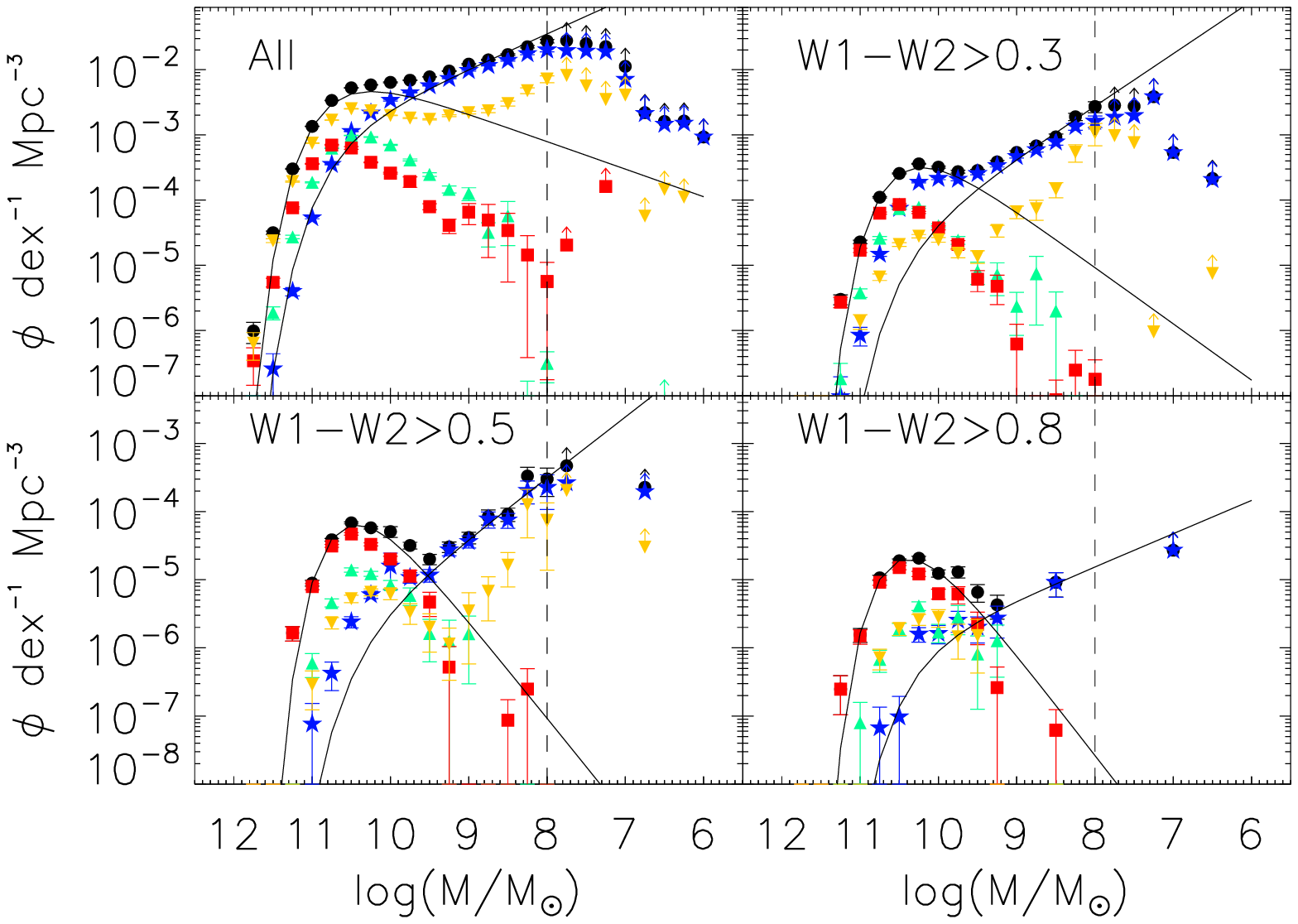}
\caption{The 1/V$_{Max}$ stellar mass function points for the full sample (top
left), W1 - W2 $\geq$ 0.3 (top right), W1 - W2 $\geq$ 0.5 (bottom left) and 
W1 - W2 $\geq$ 0.8 (bottom right). The total stellar mass function in each colour
range is shown as black circles. Optical AGNs (broad and narrow line) are shown as
red squares, composite galaxies as green triangles, star-forming galaxies as blue stars and unclassified
galaxies as yellow inverted triangles. The vertical dashed line is the limit below which the stellar mass function becomes uncertain due to incompleteness.}
\label{Fig:MFBPT}
\end{figure*}

The classifications of galaxies in Figure \ref{Fig:MFBPT} are based on optical
emission line ratios with divisions between star-forming, composite, and AGN regions
from \citet{Kewley2001} and \citet{Kauffmann2003a}. The AGN classification
includes both broad and narrow line systems. Galaxies are unclassified if they
have S/N$<$3 for the H$\alpha$, H$\beta$,
[\ion{O}{iii} 5007] or [\ion{N}{ii} 6584] lines. As in Figure \ref{Fig:MSchechter},
points below 10$^{8}$M$_{\odot}$ are effected by surface brightness
incompleteness and are, therefore, lower limits.

Figure \ref{Fig:MFBPT} shows that optically classified AGNs are found almost
exclusively at high stellar masses, in agreement with \citet{Kauffmann2003a}.
However, the majority of high stellar mass galaxies are optically unclassified
because they lack strong emission lines. Similarly, nearly all massive galaxies
with W1-W2 $> 0.5$ and W1 - W2 $>0.8$ are strongly dominated by AGNs in
agreement with both the optically selected AGN samples of \citet{Kauffmann2003a}
and the X-ray and IR-selected samples of \citet{Xue2010}. The fraction of
optically classified AGNs drops rapidly towards lower masses. For all infrared
colours, the optical emission lines of low mass galaxies are dominated by star
formation, consistent with the optical emission line studies of \citet{Kewley2006}.

\subsection{The Impact of Metallicity and Specific Star Formation Rates on
Galaxy Colour}

\begin{figure*}
\includegraphics[width=\textwidth]{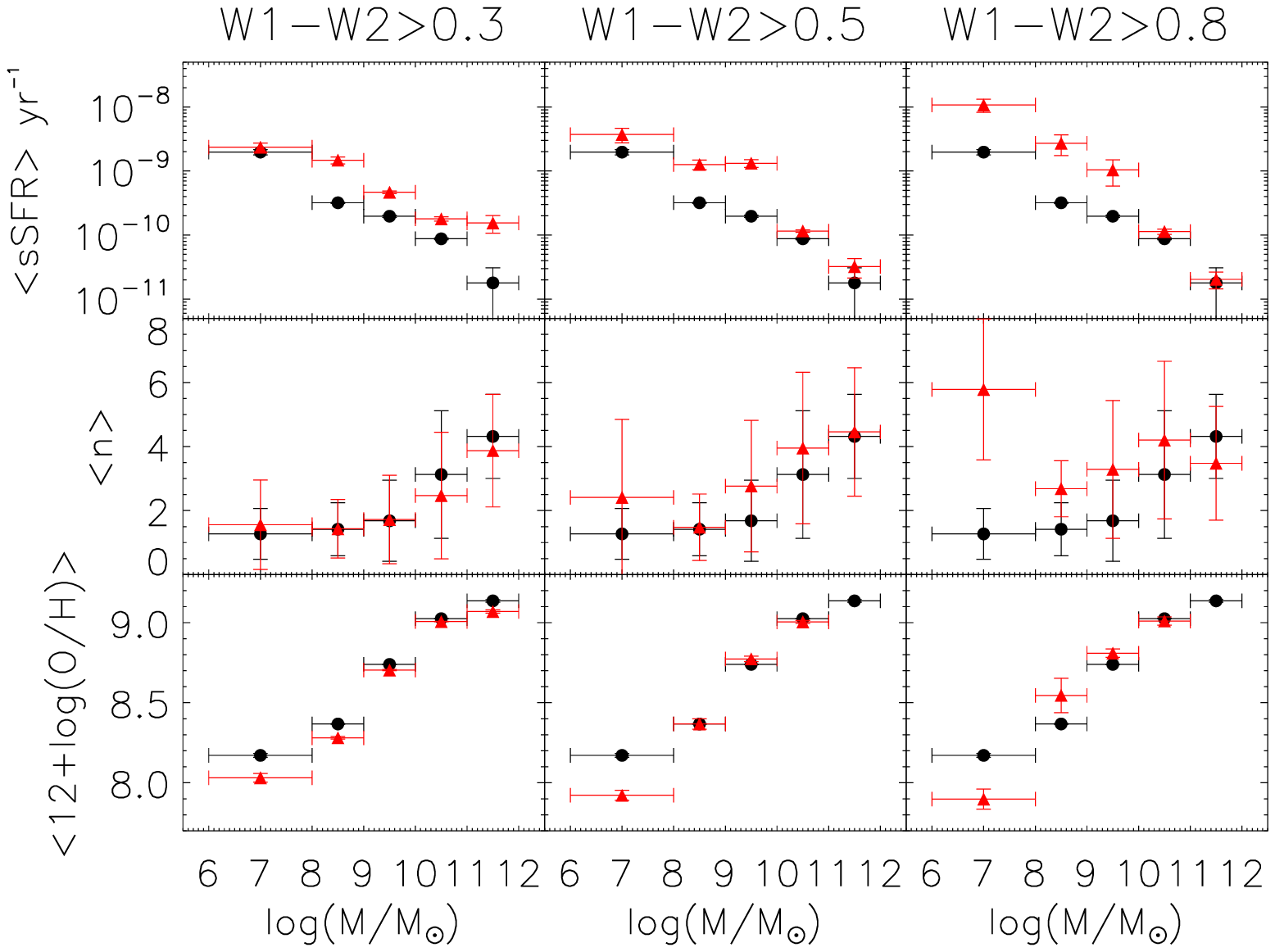}
\caption{From top to bottom, the average specific star formation rates, S\'{e}rsic 
indices and metallicities of sample galaxies as functions of stellar mass and 
WISE colour. 
Black circles are the values calculated using only galaxies with W1-W2$<$0.3
while red triangles are calculated using galaxies with W1-W2$\geq$0.3 (left),
W1-W2$\geq$0.5 (middle) and W1-W2$\geq$0.8 (right).
All averages are weighted using the inverse of each galaxy's V$_{Max}$ value.}
\label{Fig:ssfrmet}
\end{figure*}

Figure \ref{Fig:ssfrmet} shows the average specific star formation rates (sSFRs), 
S\'{e}rsic indices (n)
and metallicities for galaxies with W1 - W2 $< 0.3$ in bins of stellar mass (black
points). Note that the black points in each panel are the same. The red points
show the average sSFRs, S\'{e}rsic indices and metallicities for
galaxies with W1 - W2 $\geq$ 0.3 (left), W1 - W2 $\geq$ 0.5 (middle), and W1 - W2 $\geq$ 0.8
(right). Specific star formation rates and metallicities are drawn from the
MPA-JHU catalogue \citep{Brinchmann2004,Tremonti2004}. S\'{e}rsic indices come from Table 3 of
\citet{Simard2011}. The sSFRs are only
included for galaxies with S/N $\geq$ 3 in H$\alpha$ and metallicities are only
included for galaxies with S/N $\geq$ 3 in H$\alpha$, H$\beta$, [\ion{O}{iii} 5007]
and [\ion{O}{iii} 3727]. The values for each galaxy are weighted by 1/V$_{Max}$ to
account for its detectability. The errors are the standard deviation in the bin
divided by the square root of the number of galaxies. It should be noted that the sSFRs in low mass galaxies may be susceptible to bias due to their low masses even if there is relatively little star formation. There is also a possibility that AGN activity may falsely bolster the star formation rates of these galaxies, though the positions of low mass galaxies on the BPT diagram imply little optical input from the AGNs (if present) to the H$\alpha$ emission line fluxes.


Figure \ref{Fig:ssfrmet} indicates that galaxies with red colours in the infrared
follow a similar trend in metallicity to those with bluer colours with lower mass
galaxies exhibiting lower metallicities. There is some evidence for lower
metallicities for the lowest mass, reddest galaxies but it is only a
3.2-$\sigma$ effect.

\section{Discussion}

\subsection{Galaxy Populations as a Function of Mass}

The change in slope between high and low mass galaxies in the mass and
luminosity functions of red galaxies points to two separate populations of red
galaxies. By mass these populations are well aligned with the red sequence
(massive galaxies) and the blue cloud (low mass galaxies) as demonstrated by
\citet{Blanton2005b} and \citet{Baldry2008}. Morphologically the massive red
galaxies have larger average S\'{e}rsic indices than the lower mass systems, but the 
scatter is large within individual stellar mass bins.

Starburst galaxies and AGN can both redden galaxy colours such that W1 - W2 $>
0.5$ \citep{Jarrett2011,Wright2010}. Most (85.4\%) high mass (M/M$_{\odot}
> 10^{10}$) red (W1 - W2 $>$ 0.5) galaxies are optically classified as AGN or
composites while 85.8\% of low mass (10$^{8}$ $ < $ M/M$_{\odot}$ $
< $ 10$^{9}$) red galaxies are optically classified as star-forming systems. The
differences in optical classifications of red galaxies as a function of stellar
mass indicates that different mechanisms may be responsible for heating the dust
(only $\sim13$\% of low mass galaxies are unclassified due to low S/N emission
lines so this is not the reason for the difference).  

The difference in the optical spectral properties of the high and low mass end
of the red galaxy population can be interpreted in several different ways: (1)
The dust heating that produces the red colours is powered by different physical
processes in high and low mass galaxies -- AGNs at the high mass end and star
formation at the low mass end; (2) the same AGNs are responsible for the
dust heating over the full mass spectrum of galaxies but the AGNs are not
optically visible in low mass systems due to the dusty environments; or (3)
while AGNs are largely responsible for the dust heating in massive systems AGNs
and/or star formation can contribute for low mass galaxies. 

For the following discussion, red galaxies refers to systems
with W1-W2 $>$ 0.5. This colour cut defines the AGN region in \citet{Stern2012}
and \citet{Jarrett2011} and provides better statistics than a colour cut of
W1-W2 $>$ 0.8.

\subsection{Star Formation as a Driver of Dust Heating in Red Galaxies?}

The simplest explanation for the differences in the optical emission line
properties of high and low mass red galaxies is a difference in the physical
process responsible for heating the dust. For low mass red galaxies the most
likely heating mechanism is star formation. 

Figure \ref{Fig:ssfrmet} shows that low mass red galaxies have higher sSFRs than
their bluer counterparts while high mass red and blue galaxies have comparable
sSFRs. However, there are known correlations between star formation and AGN
activity so a correlation between star formation and red colours does not rule
out AGNs as a power source \citep[e.g.]{Netzer2007, Lutz2008, Netzer2009,
Woo2012, Rosario2012}. 

One important note on stars being responsible for the red colours is that it
must be the heating by young stars rather than the galaxies being red due to the
stellar colours themselves because stars rarely get this red. \citet{Chen2014}
observed dust-free stars in the Milky Way  and found that their colours rarely
exceed W1 - W2 = 0.25. \citet{Faherty2014} and \citet{Nikutta2014}  provide a
handful of exceptions to this rule, but the classes of stars that they 
observe (Wolf-Rayet and PAGB stars) are both rare and unlikely to dominate the
colour of an entire galaxy. Similarly, T dwarfs and later-type stars may have W1
- W2 as high as 4.2 but are too dim to dominate the
light of a galaxy \citep{Kirkpatrick2011}. Though the fraction of infrared light 
contribution may vary significantly from one galaxy to another, Thermally Pulsating 
Asymptotic Giant Branch (TP-AGB) stars' emission peaks in in the 3-4$\mu$m range and 
has much redder than average [3.6]-[4.5] colour \citep{Meidt2012,Gerke2013,Melbourne2013,Villaume2015}.

While stars themselves make a negligible contribution to these red colours,
\citet{Calzetti2010,Kennicutt1998} have shown that young, hot stars can
contribute significantly to dust heating through the reproccessing of UV
radiation. It has also been noted that redder colours from star formation might
be expected in low mass galaxies due to their lower average metallicities
\citep{Hunt2010}. These lower metallicity stars produce a harder radiation field
\citep{Lee2004,Kewley2004,Moustakas2006} due to a combination of decreased mass-loss 
rates, higher Hayashi limits and a lack of line blanketing seen in higher metallicity 
stars \citep{Levesque2010} and thus additional dust heating. However, 
Figure \ref{Fig:ssfrmet} shows that the metallicity of the red and blue low mass
galaxies is similar so the harder radiation field in low metallicity systems is
unlikely to be a key factor in which dwarf galaxies exhibit red colours. 

If dust heating by star formation drives the red infrared colours, galaxies with
strong, concentrated star formation (e.g., blue compact dwarf galaxies, BCDs)
may be more likely to have red galaxy-wide colours. To examine this possibility,
we identify the 599 galaxies in this sample classified as BCDs using the
criteria of \citet{Sanchez2008}. Galaxies that are identified 
as BCDs are more likely to also have red W1 - W2 colors. There is a 5.2 $\sigma$ 
difference between the fraction of BCDs and all sample galaxies with W1 - W2 $>$ 0.5 
and a 2.5 $\sigma$ difference for galaxies with W1 - W2 $>$ 0.8.
Corroborating the idea that star formation is important in driving the red
colours of BCDs, those with W1 - W2 $\geq$ 0.8 all have W2 - W3 $>$ 4 and only
19\% (7/37) with W1 - W1 $\geq$ 0.5 have W2 - W3 $\leq$ 4. 15.0\% and 12.8\% 
of galaxies with W1 - W2 $>$ 0.5 and W1 - W2 $>$ 0.8 in the full sample also have 
W2 - W3 $>$ 4 (5.7 and 72.7 $\sigma$ difference, respectively). High W2 - W3 colour is
more likely to come from star formation than AGN activity because the colour
comes from enhanced PAH features in the W3 band \citep{Jarrett2013}. 

Given that only a small fraction of BCDs show high W1 - W2 colours
\citep{Griffith2011,Izotov2011,Izotov2014}, concentrated star formation is
clearly not a sufficient condition for making low mass galaxies red. It is also
possible that the red colours in these BCDs are actually powered by AGN. The
spectra of 5 BCDs (out of 12) show [\ion{Ne}{v} 3426] emission. 3 of these 
also show [\ion{Fe}{v} 4227] emission
\citep{Izotov2004,Thuan2005,Izotov2012}. Both
of these high ionization lines are often associated with AGN activity and 4 out
of 5 BCDs with [\ion{Ne}{v} 3426] emission and all with
[\ion{Fe}{v} 4227] emission have W1 - W2 $>$ 0.5 \citep{Izotov2012}.

A more detailed study of dust heating by star
formation and its contribution to red colours is needed, but it is beyond the
scope of this work.
 
\subsection{AGN as the Driver of Dust Heating in Red Galaxies?}

Most (85.2\%) massive (10$^{10}$ $<$ M/M$_{\odot}$ $<$ 10$^{12}$) red (W1 -
W2 $\geq$ 0.8) galaxies in this sample are optical AGN or composites. However a growing 
number of AGNs have been identified in low mass
galaxies \citep{Barth2004, Dong2007, Greene2007, Ghosh2008, Izotov2008,
Jiang2011, Ho2012, Reines2012, Schramm2013, Maksym2014, Moran2014, Satyapal2014,
Yuan2014} including  He2-10 (1.4$\times$10$^{9}$M$_{\odot}$) \citep{Reines2011,
Reines2012, Whalen2015}, Mrk 709S (2.5$\times$10$^{9}$M$_{\odot}$)
\citep{Reines2014} and SDSS J1329+3234 (2$\times$10$^{8}$M$_{\odot}$)
\citep{Secrest2015}. \citet{Reines2013} has found 151 optical
AGNs in galaxies with 10$^{8.5} < M/M_{\odot} < 10^{9.5}$, 0.6\% of
galaxies in this mass range. This fraction of low mass galaxies is similar to
the 0.3\% of galaxies with W1 - W2 $> 0.8$. However, only 2 out of 15 (13\%) 
low mass BLAGN in \citet{Reines2013} have W1 - W2 $\geq$ 0.5 (1 with W1 - W2
$\geq$ 0.8). Inversely 11 out of 136 (8\%) low mass AGNs identified with BPT
line ratios have W1 - W2 $\geq$ 0.5 (8 have W1 - W2 $\geq$ 0.8) so even if the
galaxies with red WISE colours are AGNs, they are generally not the same ones that
are optically classified as AGNs. 
 
If AGNs are important for the dust heating in these low mass galaxies,
almost all of them would have to be deeply embedded in dust despite these
systems generally being thought to have very little dust
\citep{Cook2014,Draine2007}. \citet{Akylas2012} find that 5 - 50\% of AGN could
be Compton-thick and still be consistent with the X-ray background, a lower
percentage of the overall galaxy population than is contributed by low mass
optically star-forming galaxies with W1 - W2 $\geq$ 0.5 (85\%).

We have also examined the 2MASS \citep{Skrutskie2006} colours of our sample galaxies. Only 1.6\% of the full sample has J - K$_{s} >$ 2, which is typical of QSOs \citep{Warren2000,Hutchings2003} and only 168, 41 and 10 galaxies with W1 - W2 $\geq$ 0.3, 0.5 and 0.8 were matched with a 2MASS source in the AllWISE catalog. Three galaxies with 0.3 $<$ W1 - W2 $<$ 0.5 have J - K$_{s}$ $>$ 2, which is consistent with the 2.7$\pm$1.6 sources expected in this category. None of the galaxies with W1 - W2 $>$ 0.5 have red 2MASS colours, but this is also statistically consistent with the 1.6\% found in the full sample.

Additional observations are needed to confirm the AGN nature of low mass 
red W1 - W2 galaxies. X-ray observations with facilities such as \emph{Chandra} 
and \emph{XMM-Newton} have proven effective in uncovering AGN in low mass 
galaxies \citep{Reines2011,Reines2013,Secrest2015}, though as evidenced by 
the lack of optical emission line visibility, obscuration may be an issue. 
At z=0, the hard X-ray (2-10 keV) emission of an AGN with N$_{H}$=10$^{24}$ cm$^{-2}$ 
is suppressed by more than a factor of 10 \citep{Brightman2011a}. Fortunately, 
the higher energy (10-195 keV) emission is reduced by a factor of less than 2 
\citep{Brightman2011a}. \emph{NuSTAR} \citep{Harrison2013} is capable 
of explaining this higher energy (3-79 keV) range and has already observed 
several obscured and/or low luminosity AGN \citep{Annuar2015,Lansbury2014,Lansbury2015,Madsen2015,Ursini2015,Ricci2016}.

The launch of the \emph{James Webb Space Telescope} \citep{Gardner2006} in 2018 will 
bring about the opportunity to observe the infrared spectra of low mass IR-red galaxies. 
Though the wavelength range of \emph{JWST}'s NIRspec and MIRI instruments (0.6-28 $\mu$m) 
is shorter than that of \emph{IRS} (5-38 $\mu$m), it will still be capable of 
observing spectral features such as the 6.2 $\mu$m and 11.3 $\mu$m PAH features 
and high ionization lines like [NeV 14.32$\mu$m] and [OIV 25.89$\mu$m] used in 
AGN diagnostics developed for galaxies observed with IRS \citep{Spoon2007,Tommasin2010,HernanCaballero2011,Magdis2013,Gruppioni2016}.

\subsection{Implications for Seed Black Hole Models}

Two models currently exist for the origin of supermassive black hole seeds at
the centres of galaxies: (1) the creation of a seed through the death of a
population III star and (2) the direct collapse of a massive gas cloud
\citet[and references therein]{Volonteri2012}. The fraction of very low mass
galaxies with central black holes (occupation fraction) can provide a test of
these models, but in practice measuring the occupation fraction in low mass
galaxies is difficult to do \citep{Miller2015}.

\citet{vanWassenhove2010} use these two mechanisms to seed satellites in Milky
Way type haloes. The result of the \citet{vanWassenhove2010} study is that the
occupation fraction is significantly lower for massive seeds than it is for the
population III seeds. To compare with these models which determine the
occupation fraction as a function of velocity dispersion, $V$-band luminosities
are derived from the $r$-band luminosity and $g - r$ colour according to the
transformations of \citet{jester2005} and from those values velocity
dispersion is derived using the \emph{V}-$\sigma$ relation in
\citet{vanWassenhove2010}.

For both models the occupation fraction at $z = 0$ is one for velocity
dispersions above $\sigma$ = 50 km s$^{-1}$. At velocity dispersion of $\sigma
< \sim$32 km s$^{-1}$, which corresponds to M$_{r}$= -14.4 (just beyond the last
point of the full luminosity function), the models begin to diverge with the
population III seeds continuing to have an occupation fraction of one
while direct collapse model seeds have an occupation fraction of 0.6. At this
luminosity the fraction of optical AGN have effectively gone to zero so optical
AGN would indicate either a much lower occupation fraction than predicted in
either of these models or they predict a rapidly falling fraction of central
black holes that are active as galaxy mass decreases. 

One possible explanation for the decrease in optical AGN at lower masses is the
increase in the number of AGN that are embedded in dust. Given that these are
the masses at which there is a transition from early type galaxies that have
small amounts of gas and dust in their ISM to late type galaxies that have
significant gas and dust in their ISM, it is possible that the number of deeply
embedded AGN would increase. 

While it is unlikely that all low mass galaxies with red WISE colours are AGN
that assumption can be used to place some constraints on the population.
Galaxies with W1 - W2 $\geq$ 0.5 and W1 - W2 $\geq$ 0.8 at a luminosity of
M$_{r} \sim$ -14.4 comprise $\sim$3\% and $\sim$1\% of the total number density.
For the population III seed models only $\sim$3\% and $\sim$1\% of the black
holes would, therefore, be expected to be active in these low mass galaxies
because the occupation fraction is predicted to be one. For the massive seed
models $\sim$5\% and $\sim$2\% of the galaxies would be expected to be active
depending on the colour above which all of the galaxies possess embedded AGN. In
both cases this still predicts a much smaller active fraction than observed in
more massive galaxies \citep{Kauffmann2003a}. These numbers indicate that either
there is a problem with the black hole seed models or the fraction of active
black holes drop with galaxy mass even if there is a substantial population of
embedded AGN residing in low mass late-type galaxies. We note that mid-IR colour selection only finds AGN that dominate the bolometric luminosity of the galaxy. It is possible that there is a significant fraction of weakly accreting and optically unidentified AGNs \citep{Satyapal2007, Satyapal2008, Satyapal2009} that would not be identified through W1 - W2 colour selection.

\section{Summary}

We have calculated the $r$-band luminosity and stellar mass functions of z $<$ 0.1
galaxies from AllWISE and SDSS DR7 for the full population, galaxies with 
W1 - W2 $\geq$ 0.3, W1 - W2 $\geq$ 0.5 and W1 - W2 $\geq$ 0.8. We find:

\begin{enumerate}

\item Galaxies with colours redder than W1 - W2 = 0.5 make up 0.6\% of the galaxy
population and galaxies with W1 - W2 $\geq$ 0.8 make up 0.04\% of the galaxy
population for galaxy masses 10$^{8}$ $<$ M/M$_{\sun}$ $<$ 10$^{9}$. These are
fairly rare, but not an insignificant fraction of the galaxies in the local universe.

\item Massive galaxies (M/M$_{\sun}$ $>$ 10$^{10}$) with colours redder than W1 - W2 $\geq$
0.5 make up 1.0\% of the galaxy population and galaxies with W1 - W2 $\geq$ 0.8
make up 0.3\% of the galaxy population in that mass range. Relative to the total
galaxy population massive red galaxies are also rare in the universe.

\item Most (85.2\%) massive (M/M$_{\sun}$ $>$ 10$^{10}$) galaxies with 
W1 - W2 $\geq$ 0.8 are optically classified as AGN or composites, in agreement
with the numbers in \citet{Stern2012} and \citet{Jarrett2013} and a good indicator that
these colours are a good way to select massive galaxies with dusty AGN.

\item The physical mechanism responsible for the red colours in low mass galaxies
is less clear than it is in their higher mass counterparts. An increase in the
sSFR of red, low mass galaxies may point to star formation driving the heating.
However, the possibility exists that in at least some cases the red colours are
indicative of dust enshrouded AGN.

\item Even in the unlikely case that all of the low mass red galaxies possess
dust enshrouded AGN, both pop III and massive seed models indicate that the
fraction of black holes that are active in low mass galaxies is significantly
less than in more massive systems.

\end{enumerate}

\section*{Acknowledgements}

The authors would like to thank T.H. Jarrett for comments which helped improve the quality of this paper.

This study was funded by NSF grant AST-000167932.


Funding for the SDSS and SDSS-II has been provided by the Alfred P. Sloan
Foundation, the Participating Institutions, the National Science Foundation, the
U.S. Department of Energy, the National Aeronautics and Space Administration,
the Japanese Monbukagakusho, the Max Planck Society, and the Higher Education
Funding Council for England. The SDSS Web Site is http://www.sdss.org/.

The SDSS is managed by the Astrophysical Research Consortium for the
Participating Institutions. The Participating Institutions are the American
Museum of Natural History, Astrophysical Institute Potsdam, University of Basel,
University of Cambridge, Case Western Reserve University, University of Chicago,
Drexel University, Fermilab, the Institute for Advanced Study, the Japan
Participation Group, Johns Hopkins University, the Joint Institute for Nuclear
Astrophysics, the Kavli Institute for Particle Astrophysics and Cosmology, the
Korean Scientist Group, the Chinese Academy of Sciences (LAMOST), Los Alamos
National Laboratory, the Max-Planck-Institute for Astronomy (MPIA), the
Max-Planck-Institute for Astrophysics (MPA), New Mexico State University, Ohio
State University, University of Pittsburgh, University of Portsmouth, Princeton
University, the United States Naval Observatory, and the University of
Washington.

This publication makes use of data products from the Wide-field Infrared Survey
Explorer, which is a joint project of the University of California, Los Angeles,
and the Jet Propulsion Laboratory/California Institute of Technology, funded by
the National Aeronautics and Space Administration.

This research has made use of the VizieR catalogue access tool, CDS Strasbourg, France. 
the original description of the VizieR service was published in A\&AS 143, 23.

This research has made use of the NASA/IPAC Extragalactic Database (NED) which is operated 
by the Jet Propulsion Laboratory, California Institute of Technology, under contract 
with the National Aeronautics and Space Administration.

\bibliography{wisebib}

\begin{thebibliography}{}
\makeatletter
\relax
\def\mn@urlcharsother{\let\do\@makeother \do\$\do\&\do\#\do\^\do\_\do\%\do\~}
\def\mn@doi{\begingroup\mn@urlcharsother \@ifnextchar [ {\mn@doi@}
  {\mn@doi@[]}}
\def\mn@doi@[#1]#2{\def\@tempa{#1}\ifx\@tempa\@empty \href
  {http://dx.doi.org/#2} {doi:#2}\else \href {http://dx.doi.org/#2} {#1}\fi
  \endgroup}
\def\mn@eprint#1#2{\mn@eprint@#1:#2::\@nil}
\def\mn@eprint@arXiv#1{\href {http://arxiv.org/abs/#1} {{\tt arXiv:#1}}}
\def\mn@eprint@dblp#1{\href {http://dblp.uni-trier.de/rec/bibtex/#1.xml}
  {dblp:#1}}
\def\mn@eprint@#1:#2:#3:#4\@nil{\def\@tempa {#1}\def\@tempb {#2}\def\@tempc
  {#3}\ifx \@tempc \@empty \let \@tempc \@tempb \let \@tempb \@tempa \fi \ifx
  \@tempb \@empty \def\@tempb {arXiv}\fi \@ifundefined
  {mn@eprint@\@tempb}{\@tempb:\@tempc}{\expandafter \expandafter \csname
  mn@eprint@\@tempb\endcsname \expandafter{\@tempc}}}

\bibitem[\protect\citeauthoryear{{Abazajian} et~al.,}{{Abazajian}
  et~al.}{2009}]{Abazajian2009}
{Abazajian} K.~N.,  et~al., 2009, \mn@doi [\apjs]
  {10.1088/0067-0049/182/2/543}, \href
  {http://adsabs.harvard.edu/abs/2009ApJS..182..543A} {182, 543}

\bibitem[\protect\citeauthoryear{{Akylas}, {Georgakakis}, {Georgantopoulos},
  {Brightman}  \& {Nandra}}{{Akylas} et~al.}{2012}]{Akylas2012}
{Akylas} A.,  {Georgakakis} A.,  {Georgantopoulos} I.,  {Brightman} M.,
  {Nandra} K.,  2012, \mn@doi [\aap] {10.1051/0004-6361/201219387}, \href
  {http://adsabs.harvard.edu/abs/2012A%26A...546A..98A} {546, A98}

\bibitem[\protect\citeauthoryear{{Alonso-Herrero} et~al.,}{{Alonso-Herrero}
  et~al.}{2006}]{AlonsoHerrero2006}
{Alonso-Herrero} A.,  et~al., 2006, \mn@doi [\apj] {10.1086/499800}, \href
  {http://adsabs.harvard.edu/abs/2006ApJ...640..167A} {640, 167}

\bibitem[\protect\citeauthoryear{{Annuar} et~al.,}{{Annuar}
  et~al.}{2015}]{Annuar2015}
{Annuar} A.,  et~al., 2015, \mn@doi [\apj] {10.1088/0004-637X/815/1/36}, \href
  {http://adsabs.harvard.edu/abs/2015ApJ...815...36A} {815, 36}

\bibitem[\protect\citeauthoryear{{Assef} et~al.,}{{Assef}
  et~al.}{2010}]{Assef2010}
{Assef} R.~J.,  et~al., 2010, \mn@doi [\apj] {10.1088/0004-637X/713/2/970},
  \href {http://adsabs.harvard.edu/abs/2010ApJ...713..970A} {713, 970}

\bibitem[\protect\citeauthoryear{{Assef} et~al.,}{{Assef}
  et~al.}{2013}]{Assef2013}
{Assef} R.~J.,  et~al., 2013, \mn@doi [\apj] {10.1088/0004-637X/772/1/26},
  \href {http://adsabs.harvard.edu/abs/2013ApJ...772...26A} {772, 26}

\bibitem[\protect\citeauthoryear{{Baldry}, {Glazebrook}  \& {Driver}}{{Baldry}
  et~al.}{2008}]{Baldry2008}
{Baldry} I.~K.,  {Glazebrook} K.,   {Driver} S.~P.,  2008, \mn@doi [\mnras]
  {10.1111/j.1365-2966.2008.13348.x}, \href
  {http://adsabs.harvard.edu/abs/2008MNRAS.388..945B} {388, 945}

\bibitem[\protect\citeauthoryear{{Baldry} et~al.,}{{Baldry}
  et~al.}{2012}]{Baldry2012}
{Baldry} I.~K.,  et~al., 2012, \mn@doi [\mnras]
  {10.1111/j.1365-2966.2012.20340.x}, \href
  {http://adsabs.harvard.edu/abs/2012MNRAS.421..621B} {421, 621}

\bibitem[\protect\citeauthoryear{{Barth}, {Ho}, {Rutledge}  \&
  {Sargent}}{{Barth} et~al.}{2004}]{Barth2004}
{Barth} A.~J.,  {Ho} L.~C.,  {Rutledge} R.~E.,   {Sargent} W.~L.~W.,  2004,
  \mn@doi [\apj] {10.1086/383302}, \href
  {http://adsabs.harvard.edu/abs/2004ApJ...607...90B} {607, 90}

\bibitem[\protect\citeauthoryear{{Bendo} et~al.,}{{Bendo}
  et~al.}{2008}]{Bendo2008}
{Bendo} G.~J.,  et~al., 2008, \mn@doi [\mnras]
  {10.1111/j.1365-2966.2008.13567.x}, \href
  {http://adsabs.harvard.edu/abs/2008MNRAS.389..629B} {389, 629}

\bibitem[\protect\citeauthoryear{{Blanton} \& {Roweis}}{{Blanton} \&
  {Roweis}}{2007}]{BlantonRoweis2007}
{Blanton} M.~R.,  {Roweis} S.,  2007, \mn@doi [\aj] {10.1086/510127}, \href
  {http://adsabs.harvard.edu/abs/2007AJ....133..734B} {133, 734}

\bibitem[\protect\citeauthoryear{{Blanton} et~al.,}{{Blanton}
  et~al.}{2005a}]{Blanton2005b}
{Blanton} M.~R.,  et~al., 2005a, \mn@doi [\aj] {10.1086/429803}, \href
  {http://adsabs.harvard.edu/abs/2005AJ....129.2562B} {129, 2562}

\bibitem[\protect\citeauthoryear{{Blanton}, {Lupton}, {Schlegel}, {Strauss},
  {Brinkmann}, {Fukugita}  \& {Loveday}}{{Blanton}
  et~al.}{2005b}]{Blanton2005a}
{Blanton} M.~R.,  {Lupton} R.~H.,  {Schlegel} D.~J.,  {Strauss} M.~A.,
  {Brinkmann} J.,  {Fukugita} M.,   {Loveday} J.,  2005b, \mn@doi [\apj]
  {10.1086/431416}, \href {http://adsabs.harvard.edu/abs/2005ApJ...631..208B}
  {631, 208}

\bibitem[\protect\citeauthoryear{{Brightman} \& {Nandra}}{{Brightman} \&
  {Nandra}}{2011}]{Brightman2011a}
{Brightman} M.,  {Nandra} K.,  2011, \mn@doi [\mnras]
  {10.1111/j.1365-2966.2011.18207.x}, \href
  {http://adsabs.harvard.edu/abs/2011MNRAS.413.1206B} {413, 1206}

\bibitem[\protect\citeauthoryear{{Brinchmann}, {Charlot}, {White}, {Tremonti},
  {Kauffmann}, {Heckman}  \& {Brinkmann}}{{Brinchmann}
  et~al.}{2004}]{Brinchmann2004}
{Brinchmann} J.,  {Charlot} S.,  {White} S.~D.~M.,  {Tremonti} C.,  {Kauffmann}
  G.,  {Heckman} T.,   {Brinkmann} J.,  2004, \mn@doi [\mnras]
  {10.1111/j.1365-2966.2004.07881.x}, \href
  {http://adsabs.harvard.edu/abs/2004MNRAS.351.1151B} {351, 1151}

\bibitem[\protect\citeauthoryear{{Bruzual} \& {Charlot}}{{Bruzual} \&
  {Charlot}}{2003}]{Bruzual2003}
{Bruzual} G.,  {Charlot} S.,  2003, \mn@doi [\mnras]
  {10.1046/j.1365-8711.2003.06897.x}, \href
  {http://adsabs.harvard.edu/abs/2003MNRAS.344.1000B} {344, 1000}

\bibitem[\protect\citeauthoryear{{Calzetti} et~al.,}{{Calzetti}
  et~al.}{2007}]{Calzetti2007}
{Calzetti} D.,  et~al., 2007, \mn@doi [\apj] {10.1086/520082}, \href
  {http://adsabs.harvard.edu/abs/2007ApJ...666..870C} {666, 870}

\bibitem[\protect\citeauthoryear{{Calzetti} et~al.,}{{Calzetti}
  et~al.}{2010}]{Calzetti2010}
{Calzetti} D.,  et~al., 2010, \mn@doi [\apj] {10.1088/0004-637X/714/2/1256},
  \href {http://adsabs.harvard.edu/abs/2010ApJ...714.1256C} {714, 1256}

\bibitem[\protect\citeauthoryear{{Charmandaris}, {Heydari-Malayeri}  \&
  {Chatzopoulos}}{{Charmandaris} et~al.}{2008}]{Charmandaris2008}
{Charmandaris} V.,  {Heydari-Malayeri} M.,   {Chatzopoulos} E.,  2008, \mn@doi
  [\aap] {10.1051/0004-6361:200809662}, \href
  {http://adsabs.harvard.edu/abs/2008A%26A...487..567C} {487, 567}

\bibitem[\protect\citeauthoryear{{Chen} et~al.,}{{Chen}
  et~al.}{2014}]{Chen2014}
{Chen} B.-Q.,  et~al., 2014, \mn@doi [\mnras] {10.1093/mnras/stu1192}, \href
  {http://adsabs.harvard.edu/abs/2014MNRAS.443.1192C} {443, 1192}

\bibitem[\protect\citeauthoryear{{Cook} et~al.,}{{Cook}
  et~al.}{2014}]{Cook2014}
{Cook} D.~O.,  et~al., 2014, \mn@doi [\mnras] {10.1093/mnras/stu1787}, \href
  {http://adsabs.harvard.edu/abs/2014MNRAS.445..899C} {445, 899}

\bibitem[\protect\citeauthoryear{{Cutri} et~al.,}{{Cutri}
  et~al.}{2013}]{Cutri2013}
{Cutri} R.~M.,  et~al., 2013, Technical report, {Explanatory Supplement to the
  AllWISE Data Release Products}

\bibitem[\protect\citeauthoryear{{Dong} et~al.,}{{Dong}
  et~al.}{2007}]{Dong2007}
{Dong} X.,  et~al., 2007, \mn@doi [\apj] {10.1086/510899}, \href
  {http://adsabs.harvard.edu/abs/2007ApJ...657..700D} {657, 700}

\bibitem[\protect\citeauthoryear{{Donley} et~al.,}{{Donley}
  et~al.}{2012}]{Donley2012}
{Donley} J.~L.,  et~al., 2012, \mn@doi [\apj] {10.1088/0004-637X/748/2/142},
  \href {http://adsabs.harvard.edu/abs/2012ApJ...748..142D} {748, 142}

\bibitem[\protect\citeauthoryear{{Draine} et~al.,}{{Draine}
  et~al.}{2007}]{Draine2007}
{Draine} B.~T.,  et~al., 2007, \mn@doi [\apj] {10.1086/518306}, \href
  {http://adsabs.harvard.edu/abs/2007ApJ...663..866D} {663, 866}

\bibitem[\protect\citeauthoryear{{Efstathiou}, {Ellis}  \&
  {Peterson}}{{Efstathiou} et~al.}{1988}]{Efstathiou1988}
{Efstathiou} G.,  {Ellis} R.~S.,   {Peterson} B.~A.,  1988, \mnras, \href
  {http://adsabs.harvard.edu/abs/1988MNRAS.232..431E} {232, 431}

\bibitem[\protect\citeauthoryear{{Faherty}, {Shara}, {Zurek}, {Kanarek}  \&
  {Moffat}}{{Faherty} et~al.}{2014}]{Faherty2014}
{Faherty} J.~K.,  {Shara} M.~M.,  {Zurek} D.,  {Kanarek} G.,   {Moffat}
  A.~F.~J.,  2014, \mn@doi [\aj] {10.1088/0004-6256/147/5/115}, \href
  {http://adsabs.harvard.edu/abs/2014AJ....147..115F} {147, 115}

\bibitem[\protect\citeauthoryear{{Gardner} et~al.,}{{Gardner}
  et~al.}{2006}]{Gardner2006}
{Gardner} J.~P.,  et~al., 2006, \mn@doi [\ssr] {10.1007/s11214-006-8315-7},
  \href {http://adsabs.harvard.edu/abs/2006SSRv..123..485G} {123, 485}

\bibitem[\protect\citeauthoryear{{Geller} \& {Huchra}}{{Geller} \&
  {Huchra}}{1989}]{Geller1989}
{Geller} M.~J.,  {Huchra} J.~P.,  1989, \mn@doi [Science]
  {10.1126/science.246.4932.897}, \href
  {http://adsabs.harvard.edu/abs/1989Sci...246..897G} {246, 897}

\bibitem[\protect\citeauthoryear{{Gerke} \& {Kochanek}}{{Gerke} \&
  {Kochanek}}{2013}]{Gerke2013}
{Gerke} J.~R.,  {Kochanek} C.~S.,  2013, \mn@doi [\apj]
  {10.1088/0004-637X/762/1/64}, \href
  {http://adsabs.harvard.edu/abs/2013ApJ...762...64G} {762, 64}

\bibitem[\protect\citeauthoryear{{Ghosh}, {Mathur}, {Fiore}  \&
  {Ferrarese}}{{Ghosh} et~al.}{2008}]{Ghosh2008}
{Ghosh} H.,  {Mathur} S.,  {Fiore} F.,   {Ferrarese} L.,  2008, \mn@doi [\apj]
  {10.1086/591508}, \href {http://adsabs.harvard.edu/abs/2008ApJ...687..216G}
  {687, 216}

\bibitem[\protect\citeauthoryear{{Gott}, {Juri{\'c}}, {Schlegel}, {Hoyle},
  {Vogeley}, {Tegmark}, {Bahcall}  \& {Brinkmann}}{{Gott}
  et~al.}{2005}]{Gott2005}
{Gott} III J.~R.,  {Juri{\'c}} M.,  {Schlegel} D.,  {Hoyle} F.,  {Vogeley} M.,
  {Tegmark} M.,  {Bahcall} N.,   {Brinkmann} J.,  2005, \mn@doi [\apj]
  {10.1086/428890}, \href {http://adsabs.harvard.edu/abs/2005ApJ...624..463G}
  {624, 463}

\bibitem[\protect\citeauthoryear{{Goulding}, {Alexander}, {Mullaney},
  {Gelbord}, {Hickox}, {Ward}  \& {Watson}}{{Goulding}
  et~al.}{2011}]{Goulding2011}
{Goulding} A.~D.,  {Alexander} D.~M.,  {Mullaney} J.~R.,  {Gelbord} J.~M.,
  {Hickox} R.~C.,  {Ward} M.,   {Watson} M.~G.,  2011, \mn@doi [\mnras]
  {10.1111/j.1365-2966.2010.17755.x}, \href
  {http://adsabs.harvard.edu/abs/2011MNRAS.411.1231G} {411, 1231}

\bibitem[\protect\citeauthoryear{{Greene} \& {Ho}}{{Greene} \&
  {Ho}}{2007}]{Greene2007}
{Greene} J.~E.,  {Ho} L.~C.,  2007, \mn@doi [\apj] {10.1086/520497}, \href
  {http://adsabs.harvard.edu/abs/2007ApJ...667..131G} {667, 131}

\bibitem[\protect\citeauthoryear{{Griffith} et~al.,}{{Griffith}
  et~al.}{2011}]{Griffith2011}
{Griffith} R.~L.,  et~al., 2011, \mn@doi [\apjl] {10.1088/2041-8205/736/1/L22},
  \href {http://adsabs.harvard.edu/abs/2011ApJ...736L..22G} {736, L22}

\bibitem[\protect\citeauthoryear{{Gruppioni} et~al.,}{{Gruppioni}
  et~al.}{2016}]{Gruppioni2016}
{Gruppioni} C.,  et~al., 2016, preprint, \href
  {http://adsabs.harvard.edu/abs/2016arXiv160302818G} {} (\mn@eprint {arXiv}
  {1603.02818})

\bibitem[\protect\citeauthoryear{{Harrison}, {Colless}, {Kuntschner}, {Couch},
  {de Propris}  \& {Pracy}}{{Harrison} et~al.}{2010}]{Harrison2010}
{Harrison} C.~D.,  {Colless} M.,  {Kuntschner} H.,  {Couch} W.~J.,  {de
  Propris} R.,   {Pracy} M.~B.,  2010, \mn@doi [\mnras]
  {10.1111/j.1365-2966.2010.17349.x}, \href
  {http://adsabs.harvard.edu/abs/2010MNRAS.409.1455H} {409, 1455}

\bibitem[\protect\citeauthoryear{{Harrison} et~al.,}{{Harrison}
  et~al.}{2013}]{Harrison2013}
{Harrison} F.~A.,  et~al., 2013, \mn@doi [\apj] {10.1088/0004-637X/770/2/103},
  \href {http://adsabs.harvard.edu/abs/2013ApJ...770..103H} {770, 103}

\bibitem[\protect\citeauthoryear{{Hern{\'a}n-Caballero} \&
  {Hatziminaoglou}}{{Hern{\'a}n-Caballero} \&
  {Hatziminaoglou}}{2011}]{HernanCaballero2011}
{Hern{\'a}n-Caballero} A.,  {Hatziminaoglou} E.,  2011, \mn@doi [\mnras]
  {10.1111/j.1365-2966.2011.18413.x}, \href
  {http://adsabs.harvard.edu/abs/2011MNRAS.414..500H} {414, 500}

\bibitem[\protect\citeauthoryear{{Ho}, {Kim}  \& {Terashima}}{{Ho}
  et~al.}{2012}]{Ho2012}
{Ho} L.~C.,  {Kim} M.,   {Terashima} Y.,  2012, \mn@doi [\apjl]
  {10.1088/2041-8205/759/1/L16}, \href
  {http://adsabs.harvard.edu/abs/2012ApJ...759L..16H} {759, L16}

\bibitem[\protect\citeauthoryear{{Hunt}, {Thuan}, {Izotov}  \&
  {Sauvage}}{{Hunt} et~al.}{2010}]{Hunt2010}
{Hunt} L.~K.,  {Thuan} T.~X.,  {Izotov} Y.~I.,   {Sauvage} M.,  2010, \mn@doi
  [\apj] {10.1088/0004-637X/712/1/164}, \href
  {http://adsabs.harvard.edu/abs/2010ApJ...712..164H} {712, 164}

\bibitem[\protect\citeauthoryear{{Hutchings}, {Maddox}, {Cutri}  \&
  {Nelson}}{{Hutchings} et~al.}{2003}]{Hutchings2003}
{Hutchings} J.~B.,  {Maddox} N.,  {Cutri} R.~M.,   {Nelson} B.~O.,  2003,
  \mn@doi [\aj] {10.1086/375650}, \href
  {http://adsabs.harvard.edu/abs/2003AJ....126...63H} {126, 63}

\bibitem[\protect\citeauthoryear{{Izotov} \& {Thuan}}{{Izotov} \&
  {Thuan}}{2008}]{Izotov2008}
{Izotov} Y.~I.,  {Thuan} T.~X.,  2008, \mn@doi [\apj] {10.1086/591660}, \href
  {http://adsabs.harvard.edu/abs/2008ApJ...687..133I} {687, 133}

\bibitem[\protect\citeauthoryear{{Izotov}, {Noeske}, {Guseva}, {Papaderos},
  {Thuan}  \& {Fricke}}{{Izotov} et~al.}{2004}]{Izotov2004}
{Izotov} Y.~I.,  {Noeske} K.~G.,  {Guseva} N.~G.,  {Papaderos} P.,  {Thuan}
  T.~X.,   {Fricke} K.~J.,  2004, \mn@doi [\aap] {10.1051/0004-6361:20040006},
  \href {http://adsabs.harvard.edu/abs/2004A%26A...415L..27I} {415, L27}

\bibitem[\protect\citeauthoryear{{Izotov}, {Guseva}, {Fricke}  \&
  {Henkel}}{{Izotov} et~al.}{2011}]{Izotov2011}
{Izotov} Y.~I.,  {Guseva} N.~G.,  {Fricke} K.~J.,   {Henkel} C.,  2011, \mn@doi
  [\aap] {10.1051/0004-6361/201118402}, \href
  {http://adsabs.harvard.edu/abs/2011A%26A...536L...7I} {536, L7}

\bibitem[\protect\citeauthoryear{{Izotov}, {Thuan}  \& {Privon}}{{Izotov}
  et~al.}{2012}]{Izotov2012}
{Izotov} Y.~I.,  {Thuan} T.~X.,   {Privon} G.,  2012, \mn@doi [\mnras]
  {10.1111/j.1365-2966.2012.22051.x}, \href
  {http://adsabs.harvard.edu/abs/2012MNRAS.427.1229I} {427, 1229}

\bibitem[\protect\citeauthoryear{{Izotov}, {Guseva}, {Fricke}  \&
  {Henkel}}{{Izotov} et~al.}{2014}]{Izotov2014}
{Izotov} Y.~I.,  {Guseva} N.~G.,  {Fricke} K.~J.,   {Henkel} C.,  2014, \mn@doi
  [\aap] {10.1051/0004-6361/201322338}, \href
  {http://adsabs.harvard.edu/abs/2014A%26A...561A..33I} {561, A33}

\bibitem[\protect\citeauthoryear{{Jarrett} et~al.,}{{Jarrett}
  et~al.}{2011}]{Jarrett2011}
{Jarrett} T.~H.,  et~al., 2011, \mn@doi [\apj] {10.1088/0004-637X/735/2/112},
  \href {http://adsabs.harvard.edu/abs/2011ApJ...735..112J} {735, 112}

\bibitem[\protect\citeauthoryear{{Jarrett} et~al.,}{{Jarrett}
  et~al.}{2013}]{Jarrett2013}
{Jarrett} T.~H.,  et~al., 2013, \mn@doi [\aj] {10.1088/0004-6256/145/1/6},
  \href {http://adsabs.harvard.edu/abs/2013AJ....145....6J} {145, 6}

\bibitem[\protect\citeauthoryear{{Jester} et~al.,}{{Jester}
  et~al.}{2005}]{jester2005}
{Jester} S.,  et~al., 2005, \mn@doi [\aj] {10.1086/432466}, \href
  {http://adsabs.harvard.edu/abs/2005AJ....130..873J} {130, 873}

\bibitem[\protect\citeauthoryear{{Jiang}, {Greene}, {Ho}, {Xiao}  \&
  {Barth}}{{Jiang} et~al.}{2011}]{Jiang2011}
{Jiang} Y.-F.,  {Greene} J.~E.,  {Ho} L.~C.,  {Xiao} T.,   {Barth} A.~J.,
  2011, \mn@doi [\apj] {10.1088/0004-637X/742/2/68}, \href
  {http://adsabs.harvard.edu/abs/2011ApJ...742...68J} {742, 68}

\bibitem[\protect\citeauthoryear{{Kauffmann} et~al.,}{{Kauffmann}
  et~al.}{2003}]{Kauffmann2003a}
{Kauffmann} G.,  et~al., 2003, \mn@doi [\mnras]
  {10.1046/j.1365-8711.2003.06292.x}, \href
  {http://adsabs.harvard.edu/abs/2003MNRAS.341...54K} {341, 54}

\bibitem[\protect\citeauthoryear{{Kennicutt}}{{Kennicutt}}{1998}]{Kennicutt1998}
{Kennicutt} Jr. R.~C.,  1998, \mn@doi [\araa] {10.1146/annurev.astro.36.1.189},
  \href {http://adsabs.harvard.edu/abs/1998ARA%26A..36..189K} {36, 189}

\bibitem[\protect\citeauthoryear{{Kewley}, {Dopita}, {Sutherland}, {Heisler}
  \& {Trevena}}{{Kewley} et~al.}{2001}]{Kewley2001}
{Kewley} L.~J.,  {Dopita} M.~A.,  {Sutherland} R.~S.,  {Heisler} C.~A.,
  {Trevena} J.,  2001, \mn@doi [\apj] {10.1086/321545}, \href
  {http://adsabs.harvard.edu/abs/2001ApJ...556..121K} {556, 121}

\bibitem[\protect\citeauthoryear{{Kewley}, {Geller}  \& {Jansen}}{{Kewley}
  et~al.}{2004}]{Kewley2004}
{Kewley} L.~J.,  {Geller} M.~J.,   {Jansen} R.~A.,  2004, \mn@doi [\aj]
  {10.1086/382723}, \href {http://adsabs.harvard.edu/abs/2004AJ....127.2002K}
  {127, 2002}

\bibitem[\protect\citeauthoryear{{Kewley}, {Groves}, {Kauffmann}  \&
  {Heckman}}{{Kewley} et~al.}{2006}]{Kewley2006}
{Kewley} L.~J.,  {Groves} B.,  {Kauffmann} G.,   {Heckman} T.,  2006, \mn@doi
  [\mnras] {10.1111/j.1365-2966.2006.10859.x}, \href
  {http://adsabs.harvard.edu/abs/2006MNRAS.372..961K} {372, 961}

\bibitem[\protect\citeauthoryear{{Kirkpatrick} et~al.,}{{Kirkpatrick}
  et~al.}{2011}]{Kirkpatrick2011}
{Kirkpatrick} J.~D.,  et~al., 2011, \mn@doi [\apjs]
  {10.1088/0067-0049/197/2/19}, \href
  {http://adsabs.harvard.edu/abs/2011ApJS..197...19K} {197, 19}

\bibitem[\protect\citeauthoryear{{Lansbury} et~al.,}{{Lansbury}
  et~al.}{2014}]{Lansbury2014}
{Lansbury} G.~B.,  et~al., 2014, \mn@doi [\apj] {10.1088/0004-637X/785/1/17},
  \href {http://adsabs.harvard.edu/abs/2014ApJ...785...17L} {785, 17}

\bibitem[\protect\citeauthoryear{{Lansbury} et~al.,}{{Lansbury}
  et~al.}{2015}]{Lansbury2015}
{Lansbury} G.~B.,  et~al., 2015, \mn@doi [\apj] {10.1088/0004-637X/809/2/115},
  \href {http://adsabs.harvard.edu/abs/2015ApJ...809..115L} {809, 115}

\bibitem[\protect\citeauthoryear{{Lee}, {Salzer}  \& {Melbourne}}{{Lee}
  et~al.}{2004}]{Lee2004}
{Lee} J.~C.,  {Salzer} J.~J.,   {Melbourne} J.,  2004, \mn@doi [\apj]
  {10.1086/425156}, \href {http://adsabs.harvard.edu/abs/2004ApJ...616..752L}
  {616, 752}

\bibitem[\protect\citeauthoryear{{Lee}, {Hwang}  \& {Ko}}{{Lee}
  et~al.}{2013}]{Lee2013}
{Lee} J.~C.,  {Hwang} H.~S.,   {Ko} J.,  2013, \mn@doi [\apj]
  {10.1088/0004-637X/774/1/62}, \href
  {http://adsabs.harvard.edu/abs/2013ApJ...774...62L} {774, 62}

\bibitem[\protect\citeauthoryear{{Levesque}, {Kewley}  \& {Larson}}{{Levesque}
  et~al.}{2010}]{Levesque2010}
{Levesque} E.~M.,  {Kewley} L.~J.,   {Larson} K.~L.,  2010, \mn@doi [\aj]
  {10.1088/0004-6256/139/2/712}, \href
  {http://adsabs.harvard.edu/abs/2010AJ....139..712L} {139, 712}

\bibitem[\protect\citeauthoryear{{Loveday}}{{Loveday}}{2000}]{Loveday2000}
{Loveday} J.,  2000, \mn@doi [\mnras] {10.1046/j.1365-8711.2000.03179.x}, \href
  {http://adsabs.harvard.edu/abs/2000MNRAS.312..557L} {312, 557}

\bibitem[\protect\citeauthoryear{{Lutz} et~al.,}{{Lutz}
  et~al.}{2008}]{Lutz2008}
{Lutz} D.,  et~al., 2008, \mn@doi [\apj] {10.1086/590367}, \href
  {http://adsabs.harvard.edu/abs/2008ApJ...684..853L} {684, 853}

\bibitem[\protect\citeauthoryear{{Madsen} et~al.,}{{Madsen}
  et~al.}{2015}]{Madsen2015}
{Madsen} K.~K.,  et~al., 2015, \mn@doi [\apj] {10.1088/0004-637X/812/1/14},
  \href {http://adsabs.harvard.edu/abs/2015ApJ...812...14M} {812, 14}

\bibitem[\protect\citeauthoryear{{Magdis} et~al.,}{{Magdis}
  et~al.}{2013}]{Magdis2013}
{Magdis} G.~E.,  et~al., 2013, \mn@doi [\aap] {10.1051/0004-6361/201322226},
  \href {http://adsabs.harvard.edu/abs/2013A%26A...558A.136M} {558, A136}

\bibitem[\protect\citeauthoryear{{Mainzer} et~al.,}{{Mainzer}
  et~al.}{2011}]{Mainzer2011}
{Mainzer} A.,  et~al., 2011, \mn@doi [\apj] {10.1088/0004-637X/731/1/53}, \href
  {http://adsabs.harvard.edu/abs/2011ApJ...731...53M} {731, 53}

\bibitem[\protect\citeauthoryear{{Maksym}, {Ulmer}, {Roth}, {Irwin}, {Dupke},
  {Ho}, {Keel}  \& {Adami}}{{Maksym} et~al.}{2014}]{Maksym2014}
{Maksym} W.~P.,  {Ulmer} M.~P.,  {Roth} K.~C.,  {Irwin} J.~A.,  {Dupke} R.,
  {Ho} L.~C.,  {Keel} W.~C.,   {Adami} C.,  2014, \mn@doi [\mnras]
  {10.1093/mnras/stu1485}, \href
  {http://adsabs.harvard.edu/abs/2014MNRAS.444..866M} {444, 866}

\bibitem[\protect\citeauthoryear{{Mateos} et~al.,}{{Mateos}
  et~al.}{2012}]{Mateos2012}
{Mateos} S.,  et~al., 2012, \mn@doi [\mnras]
  {10.1111/j.1365-2966.2012.21843.x}, \href
  {http://adsabs.harvard.edu/abs/2012MNRAS.426.3271M} {426, 3271}

\bibitem[\protect\citeauthoryear{{Mateos}, {Alonso-Herrero}, {Carrera},
  {Blain}, {Severgnini}, {Caccianiga}  \& {Ruiz}}{{Mateos}
  et~al.}{2013}]{Mateos2013}
{Mateos} S.,  {Alonso-Herrero} A.,  {Carrera} F.~J.,  {Blain} A.,  {Severgnini}
  P.,  {Caccianiga} A.,   {Ruiz} A.,  2013, \mn@doi [\mnras]
  {10.1093/mnras/stt953}, \href
  {http://adsabs.harvard.edu/abs/2013MNRAS.434..941M} {434, 941}

\bibitem[\protect\citeauthoryear{{Meidt} et~al.,}{{Meidt}
  et~al.}{2012}]{Meidt2012}
{Meidt} S.~E.,  et~al., 2012, \mn@doi [\apjl] {10.1088/2041-8205/748/2/L30},
  \href {http://adsabs.harvard.edu/abs/2012ApJ...748L..30M} {748, L30}

\bibitem[\protect\citeauthoryear{{Melbourne} \& {Boyer}}{{Melbourne} \&
  {Boyer}}{2013}]{Melbourne2013}
{Melbourne} J.,  {Boyer} M.~L.,  2013, \mn@doi [\apj]
  {10.1088/0004-637X/764/1/30}, \href
  {http://adsabs.harvard.edu/abs/2013ApJ...764...30M} {764, 30}

\bibitem[\protect\citeauthoryear{{Miller}, {Gallo}, {Greene}, {Kelly}, {Treu},
  {Woo}  \& {Baldassare}}{{Miller} et~al.}{2015}]{Miller2015}
{Miller} B.~P.,  {Gallo} E.,  {Greene} J.~E.,  {Kelly} B.~C.,  {Treu} T.,
  {Woo} J.-H.,   {Baldassare} V.,  2015, \mn@doi [\apj]
  {10.1088/0004-637X/799/1/98}, \href
  {http://adsabs.harvard.edu/abs/2015ApJ...799...98M} {799, 98}

\bibitem[\protect\citeauthoryear{{Montero-Dorta} \& {Prada}}{{Montero-Dorta} \&
  {Prada}}{2009}]{Montero2009}
{Montero-Dorta} A.~D.,  {Prada} F.,  2009, \mn@doi [\mnras]
  {10.1111/j.1365-2966.2009.15197.x}, \href
  {http://adsabs.harvard.edu/abs/2009MNRAS.399.1106M} {399, 1106}

\bibitem[\protect\citeauthoryear{{Moran}, {Shahinyan}, {Sugarman}, {V{\'e}lez}
  \& {Eracleous}}{{Moran} et~al.}{2014}]{Moran2014}
{Moran} E.~C.,  {Shahinyan} K.,  {Sugarman} H.~R.,  {V{\'e}lez} D.~O.,
  {Eracleous} M.,  2014, \mn@doi [\aj] {10.1088/0004-6256/148/6/136}, \href
  {http://adsabs.harvard.edu/abs/2014AJ....148..136M} {148, 136}

\bibitem[\protect\citeauthoryear{{Moustakas}, {Kennicutt}  \&
  {Tremonti}}{{Moustakas} et~al.}{2006}]{Moustakas2006}
{Moustakas} J.,  {Kennicutt} Jr. R.~C.,   {Tremonti} C.~A.,  2006, \mn@doi
  [\apj] {10.1086/500964}, \href
  {http://adsabs.harvard.edu/abs/2006ApJ...642..775M} {642, 775}

\bibitem[\protect\citeauthoryear{{Netzer}}{{Netzer}}{2009}]{Netzer2009}
{Netzer} H.,  2009, \mn@doi [\apj] {10.1088/0004-637X/695/1/793}, \href
  {http://adsabs.harvard.edu/abs/2009ApJ...695..793N} {695, 793}

\bibitem[\protect\citeauthoryear{{Netzer} et~al.,}{{Netzer}
  et~al.}{2007}]{Netzer2007}
{Netzer} H.,  et~al., 2007, \mn@doi [\apj] {10.1086/520716}, \href
  {http://adsabs.harvard.edu/abs/2007ApJ...666..806N} {666, 806}

\bibitem[\protect\citeauthoryear{{Nikutta}, {Hunt-Walker}, {Nenkova},
  {Ivezi{\'c}}  \& {Elitzur}}{{Nikutta} et~al.}{2014}]{Nikutta2014}
{Nikutta} R.,  {Hunt-Walker} N.,  {Nenkova} M.,  {Ivezi{\'c}} {\v Z}.,
  {Elitzur} M.,  2014, \mn@doi [\mnras] {10.1093/mnras/stu1087}, \href
  {http://adsabs.harvard.edu/abs/2014MNRAS.442.3361N} {442, 3361}

\bibitem[\protect\citeauthoryear{{Polletta} et~al.,}{{Polletta}
  et~al.}{2007}]{Polletta2007}
{Polletta} M.,  et~al., 2007, \mn@doi [\apj] {10.1086/518113}, \href
  {http://adsabs.harvard.edu/abs/2007ApJ...663...81P} {663, 81}

\bibitem[\protect\citeauthoryear{{Ranalli}, {Comastri}  \& {Setti}}{{Ranalli}
  et~al.}{2003}]{Ranalli2003}
{Ranalli} P.,  {Comastri} A.,   {Setti} G.,  2003, \mn@doi [\aap]
  {10.1051/0004-6361:20021600}, \href
  {http://adsabs.harvard.edu/abs/2003A%26A...399...39R} {399, 39}

\bibitem[\protect\citeauthoryear{{Reines} \& {Deller}}{{Reines} \&
  {Deller}}{2012}]{Reines2012}
{Reines} A.~E.,  {Deller} A.~T.,  2012, \mn@doi [\apjl]
  {10.1088/2041-8205/750/1/L24}, \href
  {http://adsabs.harvard.edu/abs/2012ApJ...750L..24R} {750, L24}

\bibitem[\protect\citeauthoryear{{Reines}, {Sivakoff}, {Johnson}  \&
  {Brogan}}{{Reines} et~al.}{2011}]{Reines2011}
{Reines} A.~E.,  {Sivakoff} G.~R.,  {Johnson} K.~E.,   {Brogan} C.~L.,  2011,
  \mn@doi [\nat] {10.1038/nature09724}, \href
  {http://adsabs.harvard.edu/abs/2011Natur.470...66R} {470, 66}

\bibitem[\protect\citeauthoryear{{Reines}, {Greene}  \& {Geha}}{{Reines}
  et~al.}{2013}]{Reines2013}
{Reines} A.~E.,  {Greene} J.~E.,   {Geha} M.,  2013, \mn@doi [\apj]
  {10.1088/0004-637X/775/2/116}, \href
  {http://adsabs.harvard.edu/abs/2013ApJ...775..116R} {775, 116}

\bibitem[\protect\citeauthoryear{{Reines}, {Plotkin}, {Russell}, {Mezcua},
  {Condon}, {Sivakoff}  \& {Johnson}}{{Reines} et~al.}{2014}]{Reines2014}
{Reines} A.~E.,  {Plotkin} R.~M.,  {Russell} T.~D.,  {Mezcua} M.,  {Condon}
  J.~J.,  {Sivakoff} G.~R.,   {Johnson} K.~E.,  2014, \mn@doi [\apjl]
  {10.1088/2041-8205/787/2/L30}, \href
  {http://adsabs.harvard.edu/abs/2014ApJ...787L..30R} {787, L30}

\bibitem[\protect\citeauthoryear{{Ricci} et~al.,}{{Ricci}
  et~al.}{2016}]{Ricci2016}
{Ricci} C.,  et~al., 2016, \mn@doi [\apj] {10.3847/0004-637X/820/1/5}, \href
  {http://adsabs.harvard.edu/abs/2016ApJ...820....5R} {820, 5}

\bibitem[\protect\citeauthoryear{{Richards} et~al.,}{{Richards}
  et~al.}{2006}]{Richards2006}
{Richards} G.~T.,  et~al., 2006, \mn@doi [\apjs] {10.1086/506525}, \href
  {http://adsabs.harvard.edu/abs/2006ApJS..166..470R} {166, 470}

\bibitem[\protect\citeauthoryear{{Rosario} et~al.,}{{Rosario}
  et~al.}{2012}]{Rosario2012}
{Rosario} D.~J.,  et~al., 2012, \mn@doi [\aap] {10.1051/0004-6361/201219258},
  \href {http://adsabs.harvard.edu/abs/2012A%26A...545A..45R} {545, A45}

\bibitem[\protect\citeauthoryear{{Rovilos} et~al.,}{{Rovilos}
  et~al.}{2014}]{Rovilos2014}
{Rovilos} E.,  et~al., 2014, \mn@doi [\mnras] {10.1093/mnras/stt2228}, \href
  {http://adsabs.harvard.edu/abs/2014MNRAS.438..494R} {438, 494}

\bibitem[\protect\citeauthoryear{{S{\'a}nchez Almeida},
  {Mu{\~n}oz-Tu{\~n}{\'o}n}, {Amor{\'{\i}}n}, {Aguerri}, {S{\'a}nchez-Janssen}
  \& {Tenorio-Tagle}}{{S{\'a}nchez Almeida} et~al.}{2008}]{Sanchez2008}
{S{\'a}nchez Almeida} J.,  {Mu{\~n}oz-Tu{\~n}{\'o}n} C.,  {Amor{\'{\i}}n} R.,
  {Aguerri} J.~A.,  {S{\'a}nchez-Janssen} R.,   {Tenorio-Tagle} G.,  2008,
  \mn@doi [\apj] {10.1086/590380}, \href
  {http://adsabs.harvard.edu/abs/2008ApJ...685..194S} {685, 194}

\bibitem[\protect\citeauthoryear{{Satyapal}, {Vega}, {Heckman}, {O'Halloran}
  \& {Dudik}}{{Satyapal} et~al.}{2007}]{Satyapal2007}
{Satyapal} S.,  {Vega} D.,  {Heckman} T.,  {O'Halloran} B.,   {Dudik} R.,
  2007, \mn@doi [\apjl] {10.1086/519995}, \href
  {http://adsabs.harvard.edu/abs/2007ApJ...663L...9S} {663, L9}

\bibitem[\protect\citeauthoryear{{Satyapal}, {Vega}, {Dudik}, {Abel}  \&
  {Heckman}}{{Satyapal} et~al.}{2008}]{Satyapal2008}
{Satyapal} S.,  {Vega} D.,  {Dudik} R.~P.,  {Abel} N.~P.,   {Heckman} T.,
  2008, \mn@doi [\apj] {10.1086/529014}, \href
  {http://adsabs.harvard.edu/abs/2008ApJ...677..926S} {677, 926}

\bibitem[\protect\citeauthoryear{{Satyapal}, {B{\"o}ker}, {Mcalpine},
  {Gliozzi}, {Abel}  \& {Heckman}}{{Satyapal} et~al.}{2009}]{Satyapal2009}
{Satyapal} S.,  {B{\"o}ker} T.,  {Mcalpine} W.,  {Gliozzi} M.,  {Abel} N.~P.,
  {Heckman} T.,  2009, \mn@doi [\apj] {10.1088/0004-637X/704/1/439}, \href
  {http://adsabs.harvard.edu/abs/2009ApJ...704..439S} {704, 439}

\bibitem[\protect\citeauthoryear{{Satyapal}, {Secrest}, {McAlpine}, {Ellison},
  {Fischer}  \& {Rosenberg}}{{Satyapal} et~al.}{2014}]{Satyapal2014}
{Satyapal} S.,  {Secrest} N.~J.,  {McAlpine} W.,  {Ellison} S.~L.,  {Fischer}
  J.,   {Rosenberg} J.~L.,  2014, \mn@doi [\apj] {10.1088/0004-637X/784/2/113},
  \href {http://adsabs.harvard.edu/abs/2014ApJ...784..113S} {784, 113}

\bibitem[\protect\citeauthoryear{{Schaerer} \& {de Koter}}{{Schaerer} \& {de
  Koter}}{1997}]{Schaerer1997}
{Schaerer} D.,  {de Koter} A.,  1997, \aap, \href
  {http://adsabs.harvard.edu/abs/1997A%26A...322..598S} {322, 598}

\bibitem[\protect\citeauthoryear{{Schlegel}, {Finkbeiner}  \&
  {Davis}}{{Schlegel} et~al.}{1998}]{Schlegel1998}
{Schlegel} D.~J.,  {Finkbeiner} D.~P.,   {Davis} M.,  1998, \mn@doi [\apj]
  {10.1086/305772}, \href {http://adsabs.harvard.edu/abs/1998ApJ...500..525S}
  {500, 525}

\bibitem[\protect\citeauthoryear{{Schmidt}}{{Schmidt}}{1968}]{Schmidt1968}
{Schmidt} M.,  1968, \mn@doi [\apj] {10.1086/149446}, \href
  {http://adsabs.harvard.edu/abs/1968ApJ...151..393S} {151, 393}

\bibitem[\protect\citeauthoryear{{Schramm} et~al.,}{{Schramm}
  et~al.}{2013}]{Schramm2013}
{Schramm} M.,  et~al., 2013, \mn@doi [\apj] {10.1088/0004-637X/773/2/150},
  \href {http://adsabs.harvard.edu/abs/2013ApJ...773..150S} {773, 150}

\bibitem[\protect\citeauthoryear{{Secrest} et~al.,}{{Secrest}
  et~al.}{2015}]{Secrest2015}
{Secrest} N.~J.,  et~al., 2015, \mn@doi [\apj] {10.1088/0004-637X/798/1/38},
  \href {http://adsabs.harvard.edu/abs/2015ApJ...798...38S} {798, 38}

\bibitem[\protect\citeauthoryear{{Simard}, {Mendel}, {Patton}, {Ellison}  \&
  {McConnachie}}{{Simard} et~al.}{2011}]{Simard2011}
{Simard} L.,  {Mendel} J.~T.,  {Patton} D.~R.,  {Ellison} S.~L.,
  {McConnachie} A.~W.,  2011, \mn@doi [\apjs] {10.1088/0067-0049/196/1/11},
  \href {http://adsabs.harvard.edu/abs/2011ApJS..196...11S} {196, 11}

\bibitem[\protect\citeauthoryear{{Skrutskie} et~al.,}{{Skrutskie}
  et~al.}{2006}]{Skrutskie2006}
{Skrutskie} M.~F.,  et~al., 2006, \mn@doi [\aj] {10.1086/498708}, \href
  {http://adsabs.harvard.edu/abs/2006AJ....131.1163S} {131, 1163}

\bibitem[\protect\citeauthoryear{{Spoon}, {Marshall}, {Houck}, {Elitzur},
  {Hao}, {Armus}, {Brandl}  \& {Charmandaris}}{{Spoon}
  et~al.}{2007}]{Spoon2007}
{Spoon} H.~W.~W.,  {Marshall} J.~A.,  {Houck} J.~R.,  {Elitzur} M.,  {Hao} L.,
  {Armus} L.,  {Brandl} B.~R.,   {Charmandaris} V.,  2007, \mn@doi [\apjl]
  {10.1086/511268}, \href {http://adsabs.harvard.edu/abs/2007ApJ...654L..49S}
  {654, L49}

\bibitem[\protect\citeauthoryear{{Stern} et~al.,}{{Stern}
  et~al.}{2012}]{Stern2012}
{Stern} D.,  et~al., 2012, \mn@doi [\apj] {10.1088/0004-637X/753/1/30}, \href
  {http://adsabs.harvard.edu/abs/2012ApJ...753...30S} {753, 30}

\bibitem[\protect\citeauthoryear{{Stern} et~al.,}{{Stern}
  et~al.}{2014}]{Stern2014}
{Stern} D.,  et~al., 2014, \mn@doi [\apj] {10.1088/0004-637X/794/2/102}, \href
  {http://adsabs.harvard.edu/abs/2014ApJ...794..102S} {794, 102}

\bibitem[\protect\citeauthoryear{{Strauss} et~al.,}{{Strauss}
  et~al.}{2002}]{Strauss2002}
{Strauss} M.~A.,  et~al., 2002, \mn@doi [\aj] {10.1086/342343}, \href
  {http://adsabs.harvard.edu/abs/2002AJ....124.1810S} {124, 1810}

\bibitem[\protect\citeauthoryear{{Thuan} \& {Izotov}}{{Thuan} \&
  {Izotov}}{2005}]{Thuan2005}
{Thuan} T.~X.,  {Izotov} Y.~I.,  2005, \mn@doi [\apjs] {10.1086/491657}, \href
  {http://adsabs.harvard.edu/abs/2005ApJS..161..240T} {161, 240}

\bibitem[\protect\citeauthoryear{{Tommasin}, {Spinoglio}, {Malkan}  \&
  {Fazio}}{{Tommasin} et~al.}{2010}]{Tommasin2010}
{Tommasin} S.,  {Spinoglio} L.,  {Malkan} M.~A.,   {Fazio} G.,  2010, \mn@doi
  [\apj] {10.1088/0004-637X/709/2/1257}, \href
  {http://adsabs.harvard.edu/abs/2010ApJ...709.1257T} {709, 1257}

\bibitem[\protect\citeauthoryear{{Tremonti} et~al.,}{{Tremonti}
  et~al.}{2004}]{Tremonti2004}
{Tremonti} C.~A.,  et~al., 2004, \mn@doi [\apj] {10.1086/423264}, \href
  {http://adsabs.harvard.edu/abs/2004ApJ...613..898T} {613, 898}

\bibitem[\protect\citeauthoryear{{Ursini} et~al.,}{{Ursini}
  et~al.}{2015}]{Ursini2015}
{Ursini} F.,  et~al., 2015, \mn@doi [\mnras] {10.1093/mnras/stv1527}, \href
  {http://adsabs.harvard.edu/abs/2015MNRAS.452.3266U} {452, 3266}

\bibitem[\protect\citeauthoryear{{Villaume}, {Conroy}  \& {Johnson}}{{Villaume}
  et~al.}{2015}]{Villaume2015}
{Villaume} A.,  {Conroy} C.,   {Johnson} B.~D.,  2015, \mn@doi [\apj]
  {10.1088/0004-637X/806/1/82}, \href
  {http://adsabs.harvard.edu/abs/2015ApJ...806...82V} {806, 82}

\bibitem[\protect\citeauthoryear{{Volonteri}}{{Volonteri}}{2012}]{Volonteri2012}
{Volonteri} M.,  2012, \mn@doi [Science] {10.1126/science.1220843}, \href
  {http://adsabs.harvard.edu/abs/2012Sci...337..544V} {337, 544}

\bibitem[\protect\citeauthoryear{{Warren}, {Hewett}  \& {Foltz}}{{Warren}
  et~al.}{2000}]{Warren2000}
{Warren} S.~J.,  {Hewett} P.~C.,   {Foltz} C.~B.,  2000, \mn@doi [\mnras]
  {10.1046/j.1365-8711.2000.03206.x}, \href
  {http://adsabs.harvard.edu/abs/2000MNRAS.312..827W} {312, 827}

\bibitem[\protect\citeauthoryear{{Whalen}, {Hickox}, {Reines}, {Greene},
  {Sivakoff}, {Johnson}, {Alexander}  \& {Goulding}}{{Whalen}
  et~al.}{2015}]{Whalen2015}
{Whalen} T.~J.,  {Hickox} R.~C.,  {Reines} A.~E.,  {Greene} J.~E.,  {Sivakoff}
  G.~R.,  {Johnson} K.~E.,  {Alexander} D.~M.,   {Goulding} A.~D.,  2015,
  \mn@doi [\apj] {10.1088/0004-637X/806/1/37}, \href
  {http://adsabs.harvard.edu/abs/2015ApJ...806...37W} {806, 37}

\bibitem[\protect\citeauthoryear{{Woo}, {Kim}, {Imanishi}  \& {Park}}{{Woo}
  et~al.}{2012}]{Woo2012}
{Woo} J.-H.,  {Kim} J.~H.,  {Imanishi} M.,   {Park} D.,  2012, \mn@doi [\aj]
  {10.1088/0004-6256/143/2/49}, \href
  {http://adsabs.harvard.edu/abs/2012AJ....143...49W} {143, 49}

\bibitem[\protect\citeauthoryear{{Wright} et~al.,}{{Wright}
  et~al.}{2010}]{Wright2010}
{Wright} E.~L.,  et~al., 2010, \mn@doi [\aj] {10.1088/0004-6256/140/6/1868},
  \href {http://adsabs.harvard.edu/abs/2010AJ....140.1868W} {140, 1868}

\bibitem[\protect\citeauthoryear{{Xue} et~al.,}{{Xue} et~al.}{2010}]{Xue2010}
{Xue} Y.~Q.,  et~al., 2010, \mn@doi [\apj] {10.1088/0004-637X/720/1/368}, \href
  {http://adsabs.harvard.edu/abs/2010ApJ...720..368X} {720, 368}

\bibitem[\protect\citeauthoryear{{Yan} et~al.,}{{Yan} et~al.}{2013}]{Yan2013}
{Yan} L.,  et~al., 2013, \mn@doi [\aj] {10.1088/0004-6256/145/3/55}, \href
  {http://adsabs.harvard.edu/abs/2013AJ....145...55Y} {145, 55}

\bibitem[\protect\citeauthoryear{{Yuan}, {Zhou}, {Dou}, {Dong}, {Fan}  \&
  {Wang}}{{Yuan} et~al.}{2014}]{Yuan2014}
{Yuan} W.,  {Zhou} H.,  {Dou} L.,  {Dong} X.-B.,  {Fan} X.,   {Wang} T.-G.,
  2014, \mn@doi [\apj] {10.1088/0004-637X/782/1/55}, \href
  {http://adsabs.harvard.edu/abs/2014ApJ...782...55Y} {782, 55}

\bibitem[\protect\citeauthoryear{{van Wassenhove}, {Volonteri}, {Walker}  \&
  {Gair}}{{van Wassenhove} et~al.}{2010}]{vanWassenhove2010}
{van Wassenhove} S.,  {Volonteri} M.,  {Walker} M.~G.,   {Gair} J.~R.,  2010,
  \mn@doi [\mnras] {10.1111/j.1365-2966.2010.17189.x}, \href
  {http://adsabs.harvard.edu/abs/2010MNRAS.408.1139V} {408, 1139}

\makeatother
\end{thebibliography}


\begin{table*}
\caption{Luminosity Function Schechter Parameters}
\label{Tab:LFSchechter}
\begin{tabular}{cccccc}
\hline
Colour Range & $\phi^{*}_{1}$  & $\phi^{*}_{2}$  & M$^{*}_{r}$ & $\alpha_{1}$ & $\alpha_{2}$ \\
 & Mpc$^{-3}$ & Mpc$^{-3}$ & mag & & \\
\hline
Total & 5.659$\pm$0.007$\times10^{-3}$ & 1.949$\pm$0.010$\times10^{-3}$ & -20.753$\pm$0.013 & -0.244$\pm$0.029 & -1.438$\pm$0.013\\
W1-W2$\geq$0.3 & 4.323$\pm$0.159$\times10^{-4}$ & 9.795$\pm$1.121$\times10^{-5}$ & -19.985$\pm$0.035 & 0.463$\pm$0.078 & -1.682$\pm$0.034\\
W1-W2$\geq$0.5 & 7.150$\pm$0.362$\times10^{-5}$ & 4.929$\pm$1.422$\times10^{-6}$ & -20.354$\pm$0.066 & 0.473$\pm$0.129 & -1.911$\pm$0.084\\
W1-W2$\geq$0.8 & 2.385$\pm$0.154$\times10^{-5}$ & 5.782$\pm$6.766$\times10^{-7}$ & -20.378$\pm$0.128 & 0.217$\pm$0.226 & -1.973$\pm$0.351\\
\hline
\end{tabular}
\tablecomments{\citet{Blanton2005b}-form Schechter functions fit to our data as described in Section 3.}
\end{table*}

\begin{table*}
\caption{Stellar Mass Function Schechter Parameters}
\label{Tab:MFSchechter}
\begin{tabular}{cccccc}
\hline
Colour Range & $\phi^{*}_{1}$ & $\phi^{*}_{2}$ & log(M$^{*}$/M$_{\odot}$) & $\alpha_{1}$ & $\alpha_{2}$\\
 & Mpc$^{-3}$ & Mpc$^{-3}$ & log(M$_{\odot}$) & & \\
\hline
Total & 4.407$\pm$0.004$\times10^{-3}$ & 5.833$\pm$0.545$\times10^{-4}$ & 10.620$\pm$0.005 & -0.557$\pm$0.024 & -1.524$\pm$0.018 \\
W1-W2$\geq$0.3 & 3.733$\pm$0.066$\times10^{-4}$ & 1.757$\pm$0.408$\times10^{-5}$ & 10.281$\pm$0.015 & -0.137$\pm$0.064 & -1.804$\pm$0.057 \\
W1-W2$\geq$0.5 & 7.095$\pm$0.315$\times10^{-5}$ & 1.256$\pm$0.595$\times10^{-6}$ & 10.280$\pm$0.025 & 0.423$\pm$0.123 & -1.892$\pm$0.124 \\
W1-W2$\geq$0.8 & 2.178$\pm$0.211$\times10^{-5}$ & 5.593$\pm$7.798$\times10^{-7}$ & 10.216$\pm$0.043 & 0.479$\pm$0.224 & -1.487$\pm$0.397 \\
\hline
\end{tabular}
\tablecomments{\citet{Baldry2012}-form Schechter functions fit to our data as described in Section 3.}
\end{table*}

\bsp
\label{lastpage}
\end{document}